\documentclass[preprint,12pt]{emulateapj}
\usepackage{epsfig}
\usepackage{natbib}
\usepackage{amssymb}
\usepackage{amsbsy}
\usepackage{natbib}
\usepackage[mathcal]{euscript}
\bibpunct{(}{)}{,}{a}{}{,}
\begin{document}

\def\fraction#1/#2{\leavevmode\kern.1em
 \raise.5ex\hbox{\the\scriptfont0 #1}\kern-.1em
 /\kern-.15em\lower.25ex\hbox{\the\scriptfont0 #2}}
\def\simlt{\vcenter{\hbox{$<$}\offinterlineskip\hbox{$\sim$}}}
\def\simgt{\vcenter{\hbox{$>$}\offinterlineskip\hbox{$\sim$}}}
\def\spose#1{\hbox to 0pt{#1\hss}}
\def\etal{et al.\ }

\title{A Pulsation Search Among Young Brown Dwarfs and Very Low Mass Stars}
\author{Ann Marie Cody\altaffilmark{1,2} and Lynne A. Hillenbrand\altaffilmark{1}}
\altaffiltext{1}{California Institute of Technology, Department of Astrophysics, MC 249-17, Pasadena, 
CA 91125}
\altaffiltext{2}{Spitzer Science Center, California Institute of Technology, 1200 E California Blvd., Pasadena, CA
91125, USA}
\email{amc@ipac.caltech.edu}

\begin{abstract}

In 2005, Palla \& Baraffe proposed that brown dwarfs (BDs) and very low mass stars (VLMSs; $<$0.1 solar masses) 
may be unstable to radial oscillations during the pre-main-sequence deuterium burning phase. With 
associated periods of 1--4 hours, this potentially new class of pulsation offers 
unprecedented opportunities to probe the interiors and evolution of low-mass objects in the 1--15 
million year age range. Following up on reports of short-period variability in young clusters, we
designed a high-cadence photometric monitoring campaign to search for deuterium-burning pulsation 
among a sample of 348 BDs and VLMSs in the four young clusters $\sigma$~Orionis, Chamaeleon~I, IC~348, and
Upper Scorpius. In the resulting light curves we achieved sensitivity to periodic signals of amplitude
several millimagnitudes, on timescales from 15 minutes to two weeks. Despite the exquisite data quality,
we failed to detect any periodicities below seven hours. We conclude that D-burning pulsations are not
able to grow to observable amplitudes in the early pre-main sequence. In spite of the non-detection, we did
uncover a rich set of variability behavior-- both periodic and aperiodic-- on day to week timescales.
We present new compilations of variable sources from our sample, as well as three new candidate cluster members 
in Chamaeleon~I. 

 \end{abstract}


\section{Pulsation in Young Stars and Brown Dwarfs}

Investigation of the interiors and evolutionary states of stars has
long been a challenging task. Measurements of mass and age
rely heavily on theoretical models, many of which require further
calibration.  This situation has begun to improve dramatically
with the rise of asteroseismology. Precision
measurements of stellar oscillation frequencies and their differences can
furnish ages and masses to better than 10\%
\citep[e.g.,][]{2014arXiv1406.0652L}, and even internal rotation
profiles \citep[e.g.,][]{2014arXiv1405.0155K, 2012Natur.481...55B}.

One of the obvious limitations of asteroseismology is that it applies only to stars that are unstable to pulsation. 
Traditionally this has omitted most young and low-mass stars.
Recently, the first pre-main sequence (PMS) pulsating stars at intermediate masses have been identified and
characterized in young open clusters
\citep{2008ApJ...673.1088Z,2009ApJ...704.1710G}. These PMS
$\delta$~Scuti stars inhabit an instability strip in the
Hertzsprung-Russell diagram comprising spectral types A to F ($M\sim$1.5--3~$M_\odot$).
Observations of their pulsations have now been used to demonstrate a
relationship between oscillation frequencies and evolutionary status \citep{2014Sci...345..550Z}.

For stars less than 1.5~$M_\odot$, asteroseismology has yet to yield fruit.
Brown dwarfs (BDs) and very low mass stars (VLMSs) have been predicted
to undergo a pulsation instability fueled by central deuterium burning \citep[e.g.,][hereafter 
PB05]{2005A&A...432L..57P}. The pulsation theory involves destabilization via the epsilon
mechanism-- a conversion of nuclear energy to kinetic oscillations.
In brief, their main predictions are that deuterium burning brown dwarfs and very low
mass stars from $\sim$0.02 to 0.1$M_\odot$ will exhibit radial oscillations with periods from one to
four hours. The amplitudes are indeterminate, but the PB05 calculations
indicate that pulsations will grow exponentially with time if they are not damped in the
stellar interior by, e.g., convection. They argue that the mismatch
between the pulsation timescale and the convective overturn timescale
($\sim$weeks, apart from the surface layers where it is shorter)
rules out the damping scenario.  Empirical verification of this instability theory through 
detection of short-timescale periodic variability presents a new opportunity to probe the 
interior and surface properties of young low-mass objects.
In support of the pulsation hypothesis are several previous claims of
short-period variability with reported flux changes at the few to 
five percent level \citep[e.g.,][]{2004A&A...419..249S,2001A&A...367..218B,2003A&A...408..663Z}. 

To verify or refute these findings, we initiated a pulsation search in the $\sigma$ Orionis cluster 
\citep{2010ApJS..191..389C,2011ApJ...741....9C}, where we encountered over 100 periodic and
aperiodic variables. The conclusion of that work was that none of the light curves exhibited
behavior consistent with deuterium burning oscillations on few-hour timescales. We now
report the result of an expanded campaign in three other young star-forming
regions: IC~348, Chamaeleon~I, and the Upper Scorpius association. 
The expanded sample enables better coverage of parameter space in the H-R diagram, as well 
as improved statistical constraint on pulsation amplitudes. 
In addition to searching for short-period oscillations, we have also mined these
fields for longer timescale variability and new cluster members.

\section{Target selection and observations}

The prospect of detecting pulsation in young BDs and VLMSs is dependent on the 
existence of a suitable observational sample with characteristics in
the range of the prospective pulsators. Because the
pulsation instability strip calculated by PB05 is fairly narrow
compared to the characteristic range of luminosities observed in 1--10~Myr clusters, most very low mass 
members of a young, roughly coeval population may not be expected to
pulsate. Yet substantial uncertainties in luminosity estimates and
theoretical assumptions preclude determination of exactly which objects should display the
phenomenon. To harvest enough candidate pulsators for statistical evaluation, our survey
therefore relied on a large sample size. Selection of suitable clusters for the campaign involved two
requirements: 1) that the known young member population extend into the substellar regime where pulsation is predicted to 
occur, and 2) that the members be bright enough for high-precision photometry with telescope apertures of up 
to a few meters. This limited the cluster distance to 500~pc. The level of extinction in some 
star-forming regions further restricted the number of available targets. 

To assemble a list of pulsation candidates according to the above
criteria, we relied on the substantial populations of VLMS and BD
deuterium burning objects that have been 
identified in open clusters and star forming regions.  
Most cluster surveys to date have selected candidates based on
photometric colors and confirmed membership by identifying features of youth (e.g., strong H$\alpha$ emission or 
low-gravity lines) in follow-up spectra. However, the rich variety of phenomena in these regions, as 
traced by x-ray activity, photometric variability, accretion, and circumstellar disk signatures, provides 
alternative methods for uncovering young stellar and substellar objects.
We compiled a list of the young clusters and associations with VLMSs
and BDs; the regions chosen for study were $\sigma$
Orionis, Chamaeleon~I, Upper Scorpius and IC~348. Ultimately, our
multi-telescope monitoring program covered 
a significant fraction of the spectroscopically 
confirmed very low-mass objects in these young star-forming areas. The 1--10
Myr range in median ages of these clusters enables
testing the effect of not only mass, but also age, on our results.

To choose specific BDs and VLMSs suitable for the photometric monitoring campaign, we focused on targets with a high probability of
exhibiting pulsation, based on position relative to PB05's pulsation
instability strip on the H-R diagram. For most 
of the clusters, it was crucial to optimize the field positions to include as many pulsation 
candidates as possible. This task was initiated by searching the literature relevant to each of the 
four chosen clusters for spectral types later than M4 (corresponding to $\lesssim 0.2M_\odot$, and a 
high likelihood that they are still burning deuterium) as well as 
established membership, to rule out status as field dwarfs. The
collected spectral types and brightnesses were then used to 
produce H-R diagrams and compare the positions of known BDs and VLMSs with the 
deuterium-burning instability strip. We detail the selected
targets and the suite of observations conducted for each region below; a summary is provided in
Table~1.

\begin{deluxetable*}{ccccccccc}
\tabletypesize{\scriptsize}
\tablecolumns{9}
\tablewidth{0pt}
\tablecaption{Photometric observations comprising the pulsation search campaign}
\tablehead{
\colhead{Cluster} & \colhead{Telescope} & \colhead{Instrument} & \colhead{Field center} & \colhead{FOV} & 
\colhead{Dates} & \colhead{Duty}  & \colhead{Cadence} & \colhead{Band} \\
\colhead{} & \colhead{} & \colhead{} & \colhead{(R.A., Decl.)} & \colhead{} & \colhead{} & \colhead{cycle} 
& \colhead{(seconds)}  & \colhead{}   
}
\startdata
$\sigma$ Ori & CTIO 1.0m & Y4KCam & 5:38:00.6, -02:43:44 & 20$\arcmin\times$20$\arcmin$ & Dec. 27, 2007--Jan. 7, 2008 & 25\%  & 460 & I, R\\
Cha I  & CTIO 1.0m & Y4KCam & 11:09:51.0, -77:27:44 & 20$\arcmin\times$20$\arcmin$ & May 13--25, 2008  & 25\%  &  700  & i, r \\
USco   & CTIO 1.0m & Y4KCam & 16:11:08, -22:12:04  & 20$\arcmin\times$20$\arcmin$ & May 13--16; 21--22 2008  & 15\% &  700  & i, r \\
USco   & CTIO 1.0m & Y4KCam & 16:17:57.5, -23:45:41 & 20$\arcmin\times$20$\arcmin$ & May 23--25, 2008  & 15\% & 700  & i, r \\
USco   & P60  & (CCD) & 16:13:17.5, -19:27:00 & 12.$\arcmin$5$\times$12.$\arcmin$5 & June 1--14, 2008 & 13\%  & 330 & ip \\
IC 348 & P60  & (CCD) & 3:44:21.8, +32:05:43 &  12.$\arcmin$5$\times$12.$\arcmin$5 & Nov. 17--23, Nov. 28--29, 2008 & 18\% & 270 & ip, Cr \\
$\sigma$ Ori & CTIO 1.0m & Y4KCam & 5:39:31.1, -02:37:26 & 20$\arcmin\times$20$\arcmin$ & Dec. 14--24, 2008 & 28\% & 700 & I, R \\ 
USco   & P60  & (CCD) & 16:17:46.3, -20:54:18 & 12.$\arcmin$5$\times$12.$\arcmin$5 & May 14--30, 2009  & 13\% & 330 & ip, rp \\
$\sigma$ Ori & {\em Spitzer} & IRAC & 05:38:23.3, -02:40:29 & 5.$\arcmin$2$\times$5.$\arcmin$2 & Oct. 22--23, 2009 & 100\% & 31  & 3.6~$\mu$m \\
$\sigma$ Ori & {\em Spitzer} & IRAC & 05:38:26.4, -02:47:13 & 5.$\arcmin$2$\times$5.$\arcmin$2 & Oct. 22--23, 2009 & 100\% & 31  & 4.5~$\mu$m \\
IC 348 & {\em HST}  & WFC3  & 03:44:19.5, +32:06:20 & 162$\arcsec\times$81$\arcsec$ &  Jan. 29--Feb. 4, 2011  & 47\%, 30\%$^a$ & 400 & F814W \\
\enddata
\tablecomments{\label{obstable} We list the details of each observing
  run in the photometric pulsation search. In the last column, lower-case band 
letters refer to the Sloan (SDSS) system; where R-band (r, R or Cr) observations are listed, there were 
at most two per night, to assess general colors of objects (but not enough to study variability). 
Abbreviations are as follows: P60 is the Palomar 60-inch telescope, {\em HST} is the Hubble Space 
Telescope, WFC3 is the Wide-Field Imaging Camera 3, IRAC is the
{\em Spitzer} Infrared Array Camera. Duty cycle refers to the percentage of each day spent taking exposures (or reading out). Note: $^a$The 
two duty cycles listed for the {\em HST} run refer to that of a single orbit (images were acquired for 
46 of 97 minutes), and that of a single day (visits took place over $\sim$7 hours, and further 
observations did not resume until approximately one day later).}
\end{deluxetable*}

\subsection{$\sigma$ Orionis}

The $\sigma$~Orionis cluster was our primary target in the search for
pulsation, with observations of candidate very-low-mass members in
three fields. At approximately 3~Myr of age \citep{2008AJ....135.1616S}, it should
contain numerous stars and brown dwarfs that are still burning deuterium.
Two 20$\arcmin\times$20$\arcmin$ fields were initially selected
for monitoring with the Cerro Tololo Interamerican Observatory 1.0~m telescope
(``CTIO 1.0m'') from 2007 December 27 to 2008 January 7 and from 2008
December 14 to 24. We identified VLMSs and BDs in the cluster by
mining the works of \citet{1999ApJ...521..671B}, \citet{2001ApJ...556..830B}, \citet{2001A&A...377L...9B},
\citet{2003A&A...404..171B}, \citet{2004AN....325..705B}, \citet{2004A&A...424..857C},
\citet{2004AJ....128.2316S}, \citet{2004A&A...419..249S}, \citet{2005MNRAS.356.1583B},
\citet{2005MNRAS.356...89K}, \citet{2006A&A...446..501F}, \citet{2007A&A...470..903C},
\citet{2007ApJ...662.1067H}, \citet{2008A&A...478..667C}, \citet{2008ApJ...688..362L}, and
\citet{2009A&A...505.1115L}; a total of 153 confirmed and candidate
$\sigma$~Orionis members fell in the two CTIO fields, including 15
spectroscopically confirmed young BDs. Further details
on the selection, as well as a full list of objects and an image of the field of view, are provided 
in Cody \& Hillenbrand (2010; hereafter CH10).

We re-observed a subset of the CTIO targets in the infrared with the
{\em Spitzer} Space Telescope Warm Mission in 2009, focusing on a set of five
confirmed and two candidate BD members of $\sigma$~Orionis. Also
included in the {\em Spitzer} 3.6 and 4.5~$\mu$m fields were seven
additional higher mass ($>0.1M_\odot$) cluster members. All targets
were monitored continuously for a single 24 hour period, at a cadence
of $\sim$30~s. An extensive description of these observations is
available in Cody \& Hillenbrand (2011).

\subsection{Chamaeleon I}

\begin{figure}
\begin{center}
\includegraphics[scale=0.3]{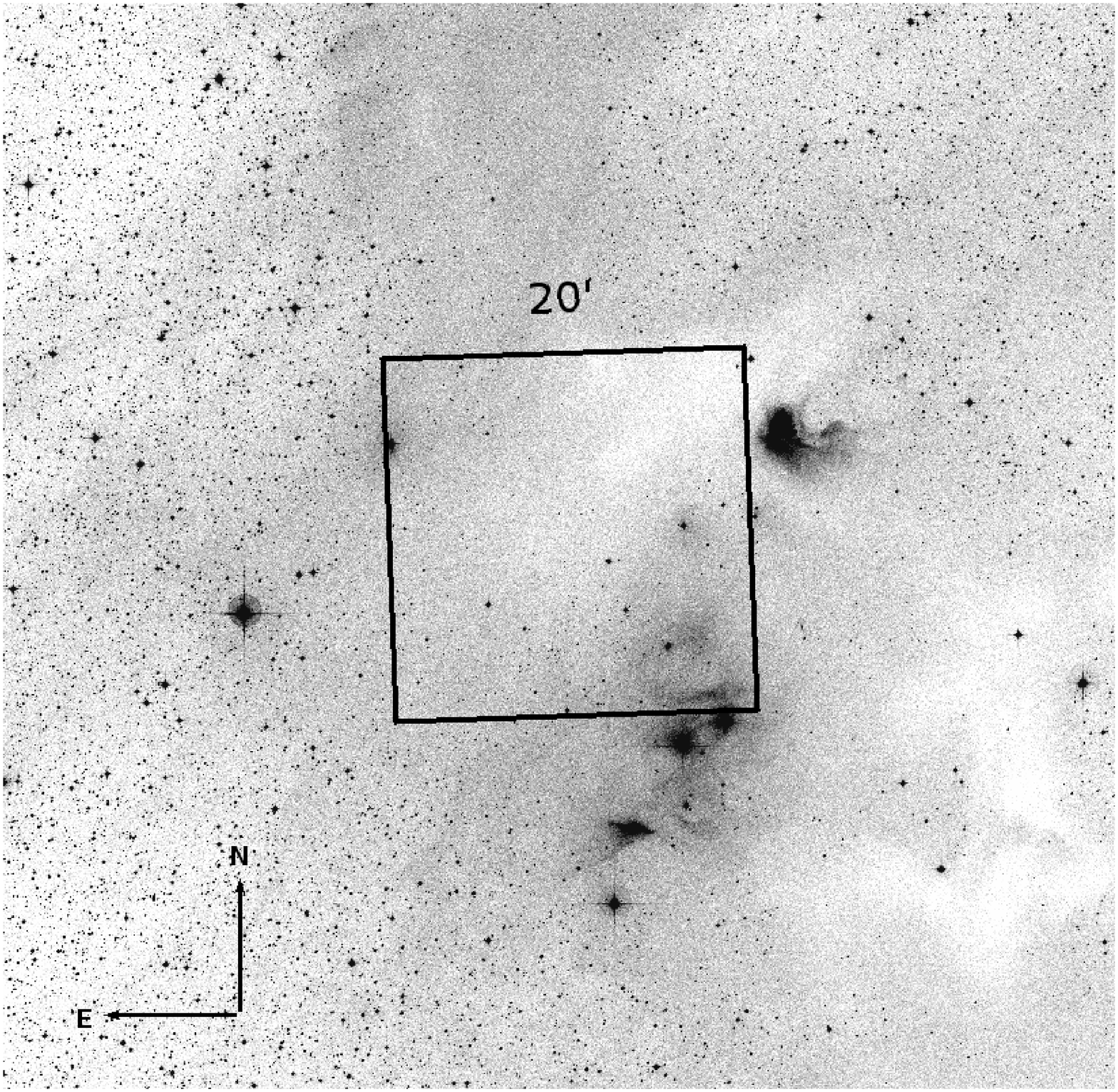}
\end{center}   
\caption[]{\label{chaIfield} The Cha~I field observed with the CTIO~1.0~m telescope is shown 
superimposed on a Digital Sky Survey image. Extinction in this region is highly variable, and we 
have avoided the most nebulous dark cloud region toward the west.}
\end{figure}

Initial target compilation for the Chamaeleon~I cluster was based on Luhman's
\citeyearpar{Luhman:2007p2372} work, plus additional members from \citet{Luhman:2008p2387},  
\citet{Luhman:2008p2378}, and \citet{Muzic:2011p8025}. These studies indicate a median age for 
the cluster of $\sim$2~Myr, making it slightly younger than $\sigma$~Orionis, and a prime
candidate for the deuterium burning pulsation search. The campaign on Cha~I involved observations 
of a single 20$\arcmin\times$20$\arcmin$ field 
with the CTIO~1.0m from May 13--25, 2008. The
field of view (FOV) was selected to maximize the number of BDs monitored and also avoid some of the dense nebulosity in this region; it is displayed in 
Figure~\ref{chaIfield}.

From the full source compilation we selected a total of 32 Cha~I
members for observation, of which 6 have spectral types consistent with substellar status 
(spectral type M6 and later) and 22 more are likely very low mass
stars with (M4 or later). The remaining four are higher mass stars and not expected to
pulsate. We have compiled the existing photometric and spectroscopic data on
the set of 32 objects, including optical through near-infrared photometry and spectral types, in 
Table~2. 

\begin{deluxetable*}{ccccccc}
\tabletypesize{\scriptsize}
\tablecolumns{7}
\tablewidth{0pt}
\tablecaption{Cha~I objects observed}
\tablehead{
\colhead{Object} & \colhead{Other ID} & \colhead{SpT} & \colhead{$i$} & \colhead{$J$} & \colhead{$H$}
 & \colhead{$K_s$} 
}
\startdata
2MASS J11105076-7718031 & ESO Halpha 568 & M4.25&  14.38 & 12.044$\pm$0.023 & 11.101$\pm$0.023 & 10.748$\pm$0.024  \\ 
2MASS J11105359-7725004 & ISO 256        & M4.5&   17.51 & 14.271$\pm$0.030 & 12.507$\pm$0.027 & 11.339$\pm$0.021  \\
2MASS J11065906-7718535 & ISO 89         & M4.25&  12.97 & 11.204$\pm$0.026 & 10.423$\pm$0.021 & 10.003$\pm$0.024  \\
2MASS J11070925-7718471 & ISO 91         & M3&     -     & 14.902$\pm$0.042 & 12.581$\pm$0.021 & 11.476$\pm$0.023  \\
2MASS J11071668-7735532 & Cha Halpha 1   & M7.75&  16.38 & 13.342$\pm$0.024 & 12.668$\pm$0.026 & 12.174$\pm$0.024  \\
2MASS J11071860-7732516 & Cha Halpha 9   & M5.5&   -     & 13.733$\pm$0.026 & 12.492$\pm$0.023 & 11.803$\pm$0.024  \\
2MASS J11072040-7729403 & ISO 99         & M4.5&   13.00 & 11.134$\pm$0.024 & 10.547$\pm$0.023 & 10.259$\pm$0.021  \\
2MASS J11073519-7734493 & CHXR 76        & M4.25&  14.39 & 12.127$\pm$0.023 & 11.279$\pm$0.023 & 10.954$\pm$0.021  \\
2MASS J11073686-7733335 & CHXR 26        & M3.5&   15.18 & 11.593$\pm$0.030 & 10.045$\pm$0.035 &  9.348$\pm$0.027  \\
2MASS J11073775-7735308 & Cha Halpha 7   & M7.75&  17.03 & 13.613$\pm$0.030 & 12.900$\pm$0.026 & 12.421$\pm$0.030  \\
2MASS J11074245-7733593 & Cha Halpha 2   & M5.25&  15.26 & 12.210$\pm$0.024 & 11.243$\pm$0.026 & 10.675$\pm$0.021  \\
2MASS J11075225-7736569 & Cha Halpha 3   & M5.5&   15.07 & 12.292$\pm$0.024 & 11.520$\pm$0.023 & 11.097$\pm$0.019  \\
2MASS J11081850-7730408 & ISO 138        & M6.5&   16.77 & 14.057$\pm$0.030 & 13.466$\pm$0.035 & 13.040$\pm$0.032  \\
  Cha J11081938-7731522 & -              & M4.75&  -     & -                & -                & -               \\
2MASS J11082238-7730277 & ISO 143        & M5&     15.51 & 12.570$\pm$0.024 & 11.651$\pm$0.027 & 11.095$\pm$0.023  \\
2MASS J11083952-7734166 & Cha Halpha 6   & M5.75&  15.06 & 12.263$\pm$0.027 & 11.479$\pm$0.024 & 11.038$\pm$0.027  \\
2MASS J11085421-7732115 & CHXR 78C       & M5.25&  15.01 & 12.310$\pm$0.026 & 11.555$\pm$0.023 & 11.224$\pm$0.024  \\
2MASS J11085596-7727132 & ISO 167        & M5.25&  17.08 & 13.514$\pm$0.031 & 12.293$\pm$0.026 & 11.619$\pm$0.025  \\
2MASS J11093543-7731390 & -              & M8.25&  -     & 15.936$\pm$0.092 & 15.022$\pm$0.087 & 14.412$\pm$0.101  \\
2MASS J11094260-7725578 & ISO 200        & M5&     15.91 & 12.329$\pm$0.027 & 11.175$\pm$0.026 & 10.552$\pm$0.028  \\
2MASS J11094742-7726290 & ESO Halpha 567 & M3.25&  16.62 & 12.767$\pm$0.027 & 11.228$\pm$0.023 & 10.236$\pm$0.022  \\
2MASS J11094918-7731197 & KG 102         & M5.5&   15.64 & 13.057$\pm$0.036 & 12.229$\pm$0.039 & 11.802$\pm$0.034  \\
2MASS J11095336-7728365 & ISO 220        & M5.75&  -     & 14.300$\pm$0.039 & 13.020$\pm$0.026 & 12.233$\pm$0.025  \\
2MASS J11100192-7725451 & LM04\_419       & M5.25&  17.46 & 13.833$\pm$0.032 & 12.605$\pm$0.026 & 12.021$\pm$0.03  \\
2MASS J11100785-7727480 & ISO 235        & M5.5&   17.79 & 13.545$\pm$0.030 & 12.097$\pm$0.026 & 11.342$\pm$0.023  \\
2MASS J11101153-7733521 & -              & M4.5&   14.24 & 12.183$\pm$0.031 & 11.192$\pm$0.023 & 10.783$\pm$0.019  \\
2MASS J11103481-7722053 & LM04\_405       & M4&     -     & 12.038$\pm$0.023 & 10.718$\pm$0.024 & 10.034$\pm$0.019  \\
2MASS J11103644-7722131 & ISO 250        & M4.75&  16.52 & 12.724$\pm$0.027 & 11.369$\pm$0.026 & 10.667$\pm$0.021  \\
2MASS J11103801-7732399 & CHXR 47        & K3&     11.90 &  9.741$\pm$0.027 &  8.687$\pm$0.047 &  8.277$\pm$0.029  \\
2MASS J11104141-7720480 & ISO 252        & M6&     17.29 & 13.860$\pm$0.030 & 12.891$\pm$0.027 & 12.266$\pm$0.023  \\
2MASS J11120288-7722483 & -              & M6&     -     & 13.588$\pm$0.030 & 12.941$\pm$0.044 & 12.510$\pm$0.030  \\
2MASS J11120351-7726009 & ISO 282        & M4.75&  -     & 13.626$\pm$0.024 & 12.587$\pm$0.025 & 11.842$\pm$0.023  \\
\enddata
\tablecomments{\label{chaIdata} Spectral types and $i$-band magnitudes are from 
\citet{Luhman:2004p2408} and \citet{Luhman:2007p2372}; $J$, $H$, and $K$ magnitudes are from 2MASS. Objects with the alternate identification LM04 are from the catalog of \citet{LopezMarti:2004p2855}.}
\end{deluxetable*}

\subsection{IC 348}

IC~348 is an appealing target in the search for pulsation, since it is
relatively compact ($<1\arcdeg$\ square), 
and its membership is very well characterized \citep{Luhman:2003p2777,Muench:2007p3882}. At a
2--3~Myr \citep{2003ApJ...593.1093L}, it is comparable in age to $\sigma$~Orionis and Cha~I.
Several previous studies have identified numerous periodic variables, which
are presumably the result of rotational modulation of spots \citep[][and references therein]{Cohen:2004p4383,Littlefair:2005p1685,Cieza:2006p1187}. The 
typical periods are near 2--3 days, but several objects have reported periods as short as 5 hours.  

\begin{figure}
\begin{center}
\includegraphics[scale=0.43]{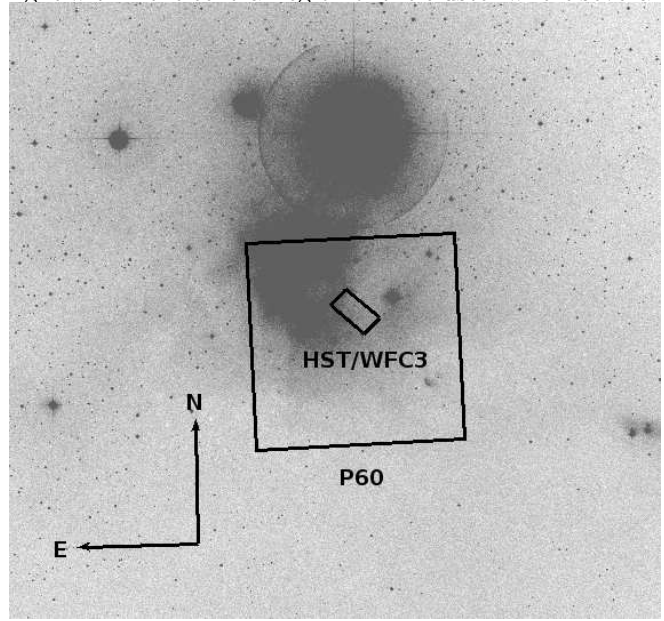}
\end{center}
\caption[]{\label{IC348field} The IC~348 fields observed with the Palomar 60-inch telescope 
(12$\arcmin$.5~$\times$~12$\arcmin$.5) and {\em HST} (81$\arcsec$~$\times$~162$\arcsec$) are shown superimposed on a 
Digital Sky Survey image. The bright B star binary o~Per lies just to the north.}
\end{figure}

We observed IC~348 from the ground with the Palomar 60-inch telescope
\citep[``P60'';][]{Cenko:2006p5614} and from space using the {\em Hubble Space Telescope} ({\em HST}) Wide Field Camera 3 
(WFC3). The $\sim$12$\arcmin$.5~$\times$~12$\arcmin$.5 ground-based field of view encompassed a significant spatial 
extent within IC~348, including the nebulous region in the cluster
center. The WFC3 field is much smaller, with a full field of view of 162~$\arcsec$~$\times$~162$\arcsec$. To maximize the 
data cadence for {\em HST} we opted to observe in the subarray mode, for which only one of two 81$\arcsec$~$\times$~162$\arcsec$ chips 
was used. Since the ground-based photometry preceded the space observations by more than three years, we 
were able to select several faint BD pulsation candidates that required photometry at the higher sensitivity 
levels afforded by {\em HST}. Therefore the WFC3 field did not cover an additional region, but rather fell 
within the previous ground-based FOV, encompassing four BDs and and two VLMSs. Two of these (L761 and L1434) do 
not have ground-based light curves since they suffered from low signal-to-noise. 
Both the ground- and space-based fields are illustrated in Figure~\ref{IC348field}.

Selection of low-mass IC~348 cluster objects was carried out by
considering the spectral types presented by \citet{Luhman:2003p2777} and
\citet{Luhman:2005p2789}.  A total of 194 members fell within
the chosen ground-based FOV. However, we did not extract photometry for stars that fell on bad pixel
columns, or for some BDs that were too faint for
adequate signal to noise. There are also a
number of brighter stars within the field for which we did not obtain
data since their point-spread functions (PSFs) were saturated and
distorted by nebulosity and scattered light within the central region of the cluster where several
bright B stars lie. For the pulsation campaign, we ultimately monitored 147 low-mass IC~348
members, including 26 BDs (spectral types M6 or later) and 65 VLMSs (M4--M6). We present a compilation 
of basic target properties in Tables~3 and 4.

\LongTables
\begin{deluxetable*}{cccccc}
\tabletypesize{\scriptsize}
\tablecolumns{6}
\tablewidth{0pt}
\tablecaption{IC~348 cluster members observed with the Palomar 60-inch
telescope}
\tablehead{
\colhead{Object} & \colhead{2MASS ID} & \colhead{$I$} & \colhead{$J$} & \colhead{$H$} &\colhead{SpT} 
}
\startdata
L13 &    2MASS J03435964+3201539 & 19.46 & 13.45 & 10.78 & M0.5  \\ 
L23 &    2MASS J03443871+3208420 & 13.97 & 11.19 &  9.97 & K3    \\ 
L26 &    2MASS J03435602+3202132 & 15.62 & 12.29 & 10.56 & K7    \\ 
L31 &    2MASS J03441816+3204570 & 15.37 & 12.09 & 10.54 & G1    \\ 
L32 &    2MASS J03443788+3208041 & 14.18 & 11.69 & 10.48 & K7    \\ 
L35 &    2MASS J03443924+3207355 & 13.21 & 10.83 &  9.95 & K3    \\ 
L37 &    2MASS J03443798+3203296 & 13.18 & 11.45 & 10.44 & K6    \\ 
L40 &    2MASS J03442972+3210398 & 14.10 & 11.93 & 10.76 & K8    \\ 
L41 &    2MASS J03442161+3210376 & 14.99 & 12.49 & 11.28 & K7    \\ 
L46 &    2MASS J03441162+3203131 & 16.24 & 12.78 & 11.22 & G8    \\ 
L48 &    2MASS J03443487+3206337 & 13.45 & 11.50 & 10.60 & K5.5  \\ 
L49 &    2MASS J03435759+3201373 & 19.63 & 14.56 & 11.89 & M0.5  \\ 
L51 &    2MASS J03441297+3201354 & 19.56 & 15.09 & 12.43 & -     \\ 
L52 &    2MASS J03444351+3207427 & 14.98 & 12.12 & 10.89 & M1    \\ 
L55 &    2MASS J03443137+3200140 & 18.06 & 13.63 & 11.65 & M0.5  \\ 
L56 &    2MASS J03440499+3209537 & 13.02 & 11.55 & 10.71 & K3.5  \\ 
L58 &    2MASS J03443854+3208006 & 14.24 & 11.94 & 10.90 & M1.25 \\ 
L61 &    2MASS J03442228+3205427 & 15.23 & 12.54 & 11.27 & K8    \\ 
L65 &    2MASS J03443398+3208541 & 13.69 & 11.85 & 10.98 & M0    \\ 
L66 &    2MASS J03442847+3207224 & 13.53 & 11.67 & 10.85 & K6.5  \\ 
L68 &    2MASS J03442851+3159539 & 14.16 & 12.00 & 11.13 & M3.5  \\ 
L69 &    2MASS J03442702+3204436 & 13.69 & 11.95 & 11.14 & M1    \\ 
L71 &    2MASS J03443257+3208558 & 14.32 & 12.11 & 11.13 & M3    \\ 
L72 &    2MASS J03442257+3201536 & 14.31 & 12.12 & 11.15 & M2.5  \\ 
L74 &    2MASS J03443426+3210497 & 14.36 & 12.14 & 11.13 & M2    \\ 
L75 &    2MASS J03444376+3210304 & 14.26 & 12.75 & 11.60 & M1.25 \\ 
L82 &    2MASS J03443740+3206118 & 13.89 & 12.09 & 11.15 & K7    \\ 
L83 &    2MASS J03443741+3209009 & 14.93 & 12.49 & 11.44 & M1    \\ 
L91 &    2MASS J03443919+3209448 & 14.76 & 12.59 & 11.52 & M2    \\ 
L92 &    2MASS J03442366+3206465 & 14.20 & 12.24 & 11.37 & M2.5  \\ 
L97 &    2MASS J03442554+3206171 & 15.98 & 12.82 & 11.59 & M2.25 \\ 
L98 &    2MASS J03443860+3205064 & 14.88 & 12.47 & 11.52 & M4    \\ 
L99 &           -                & 14.78 & 12.89 & 11.90 & M3.75 \\ 
L103 &   2MASS J03444458+3208125 & 15.73 & 12.89 & 11.87 & M2    \\ 
L105 &   2MASS J03441125+3206121 & 14.32 & 12.39 & 11.47 & M0    \\ 
L108 &   2MASS J03443869+3208567 & 14.46 & 12.49 & 11.57 & M3.25 \\ 
L115 &   2MASS J03442999+3209210 & 17.18 & 13.58 & 12.02 & M2.5  \\ 
L116 &   2MASS J03442155+3210174 & 14.57 & 12.66 & 11.70 & M1.5  \\ 
L119 &   2MASS J03442125+3205024 & 15.19 & 12.80 & 11.84 & M2.5  \\ 
L123 &   2MASS J03442457+3203571 & 15.36 & 12.85 & 11.81 & M1    \\ 
L124 &   2MASS J03435463+3200298 & 14.90 & 12.57 & 11.73 & M4.25 \\ 
L125 &   2MASS J03442166+3206248 & 14.56 & 12.52 & 11.59 & M2.75 \\ 
L128 &   2MASS J03442017+3208565 & 14.84 & 12.73 & 11.83 & M2    \\ 
L140 &   2MASS J03443568+3203035 & 15.78 & 13.47 & 12.31 & M3.25 \\ 
L142 &   2MASS J03435619+3208362 & 14.65 & 12.63 & 11.73 & M0    \\ 
L145 &   2MASS J03444129+3210252 & 14.69 & 12.65 & 11.80 & M4.75 \\ 
L146 &   2MASS J03444261+3206194 & 13.99 & 12.55 & 11.74 & M1    \\ 
L149 &   2MASS J03443698+3208342 & 15.66 & 13.07 & 12.10 & M4.75 \\ 
L153 &   2MASS J03444276+3208337 & 15.95 & 13.21 & 12.22 & M4.75 \\ 
L156 &   2MASS J03440678+3207540 & 15.31 & 13.00 & 12.12 & M4.25 \\ 
L158 &   2MASS J03444016+3209129 & 16.50 & 13.36 & 12.25 & M5    \\ 
L159 &   2MASS J03444760+3210555 & 16.60 & 13.57 & 12.29 & M4.25 \\ 
L160 &   2MASS J03440257+3201348 & 14.87 & 12.74 & 12.03 & M4.75 \\ 
L163 &   2MASS J03441122+3208161 & 15.12 & 12.78 & 12.07 & M5.25 \\ 
L165 &   2MASS J03443545+3208563 & 16.15 & 13.28 & 12.33 & M5.25 \\ 
L166 &   2MASS J03444256+3210025 & 16.85 & 13.65 & 12.43 & M4.25 \\ 
L167 &   2MASS J03444116+3210100 & 16.71 & 14.04 & 12.62 & M3    \\ 
L168 &   2MASS J03443134+3210469 & 15.84 & 13.52 & 12.40 & M4.25 \\ 
L169 &   2MASS J03441776+3204476 & 15.78 & 13.15 & 12.28 & M5.25 \\ 
L174 &   2MASS J03440410+3207170 & 15.01 & 13.02 & 12.13 & M1.5  \\ 
L182 &   2MASS J03441820+3209593 & 15.74 & 13.21 & 12.30 & M4.25 \\ 
L187 &   2MASS J03440613+3207070 & 16.36 & 13.31 & 12.42 & M4.25 \\ 
L190 &   2MASS J03442922+3201157 & 17.93 & 14.33 & 12.86 & M3.75 \\ 
L192 &   2MASS J03442364+3201526 & 18.54 & 14.47 & 12.97 & M4.5  \\ 
L194 &   2MASS J03442724+3210373 & 15.88 & 13.74 & 12.66 & M4.75 \\ 
L198 &   2MASS J03443444+3206250 & 16.07 & 13.38 & 12.54 & M5.5  \\ 
L199 &   2MASS J03435721+3201337 & - & - & - & M6.5  \\ 
L203 &   2MASS J03441810+3210534 & 18.24 & 16.04 & 13.90 & M0.75 \\ 
L205 &   2MASS J03442980+3200545 & 16.46 & 13.58 & 12.82 & M6    \\ 
L207 &   2MASS J03443030+3207426 & 17.16 & 14.01 & 12.70 & M3.5  \\ 
L210 &   2MASS J03442001+3206455 & 15.81 & 13.52 & 12.59 & M3.5  \\ 
L215 &   2MASS J03442894+3201378 & - & - & - & M3.25 \\ 
L217 &   2MASS J03444303+3210151 & 16.07 & 13.54 & 12.64 & M5    \\ 
L221 &   2MASS J03444024+3209331 & 16.57 & 14.11 & 13.03 & M4.5  \\ 
L228 &       -    		 & 18.28 & 15.07 & 13.43 & M0.5  \\ 
L230 &   2MASS J03443551+3208046 & 16.30 & 13.66 & 12.78 & M5.25 \\ 
L234 &   2MASS J03444520+3201197 & - & - & - & M5.75 \\ 
L237 &   2MASS J03442356+3209338 & 15.74 & 13.56 & 12.76 & M5    \\ 
L243 &   2MASS J03440770+3205050 & 16.71 & 14.01 & 12.98 & M4.5  \\ 
L252 &   2MASS J03442912+3207573 & 15.79 & 13.70 & 12.88 & M4.5  \\ 
L253 &   2MASS J03443165+3206534 & 16.10 & 13.58 & 12.82 & M5.5  \\ 
L254 &   2MASS J03435379+3207303 & 16.07 & 13.71 & 12.87 & M4.25 \\ 
L255 &   2MASS J03443569+3204527 & 16.10 & 13.70 & 13.01 & M5.75 \\ 
L256 &   2MASS J03435526+3207533 & 16.08 & 13.61 & 12.99 & M5.75 \\ 
L259 &   2MASS J03440362+3202341 & 16.44 & 13.54 & 12.88 & M5   \\ 
L266 &       -    		 & 16.04 & 13.73 & 12.93 & M4.75 \\ 
L276 &   2MASS J03440920+3202376 & 19.06 & 14.97 & 13.55 & M0    \\ 
L277 &   2MASS J03443943+3210081 & 16.06 & 13.91 & 13.10 & M5    \\ 
L278 &   2MASS J03443103+3205460 & 16.75 & 14.03 & 13.18 & M5.5  \\ 
L287 &   2MASS J03444111+3208073 & 17.97 & 14.59 & 13.45 & M5.25 \\ 
L298 &   2MASS J03443886+3206364 & 16.60 & 13.98 & 13.26 & M6    \\ 
L300 &   2MASS J03443896+3203196 & 16.40 & 14.11 & 13.35 & M5    \\ 
L301 &   2MASS J03442270+3201423 & 18.70 & 15.15 & 13.80 & M4.75 \\ 
L302 &   2MASS J03442027+3205437 & 17.04 & 14.24 & 13.32 & M4.75 \\ 
L303 &   2MASS J03440442+3204539 & 16.60 & 14.06 & 13.38 & M5.75 \\ 
L308 &   2MASS J03442122+3201144 & 21.03 & 16.18 & 14.24 & M4    \\ 
L312 &   2MASS J03435508+3207145 & 16.80 & 14.12 & 13.44 & M6    \\ 
L314 &   2MASS J03442256+3201277 & 18.80 & 15.13 & 13.80 & M5    \\ 
L322 &   2MASS J03441959+3202247 & 17.53 & 14.74 & 13.70 & M4.25 \\ 
L324 &   2MASS J03444522+3210557 & 17.14 & 14.56 & 13.65 & M5.75 \\ 
L325 &   2MASS J03443005+3208489 & 17.55 & 14.63 & 13.75 & M6    \\ 
L329 &   2MASS J03441558+3209218 & 17.64 & 14.57 & 13.85 & M7.5  \\ 
L334 &   2MASS J03442666+3202363 & 16.88 & 14.42 & 13.69 & M5.75 \\ 
L335 &   2MASS J03444423+3208473 & 17.34 & 14.56 & 13.76 & M5.75 \\ 
L336 &   2MASS J03443237+3203274 & 17.63 & 14.86 & 14.02 & M5.5  \\ 
L342 &   2MASS J03444130+3204534 & 17.02 & 14.49 & 13.66 & M5    \\ 
L350 &   2MASS J03441918+3205599 & 16.91 & 14.32 & 13.60 & M5.75 \\ 
L351 &   2MASS J03442575+3209059 & 17.62 & 14.69 & 13.76 & M5.5  \\ 
L353 &   2MASS J03443814+3210215 & 16.87 & 14.46 & 13.70 & M6    \\ 
L355 &   2MASS J03443920+3208136 & 18.17 & 14.88 & 14.03 & M8    \\ 
L358 &   2MASS J03441276+3210552 & 16.79 & 14.61 & 13.92 & M5.5  \\ 
L360 &   2MASS J03444371+3210479 & 16.40 & 14.54 & 13.84 & M4.75 \\ 
L363 &   2MASS J03441726+3200152 & 17.97 & 14.92 & 14.16 & M8    \\ 
L365 &   2MASS J03441022+3207344 & 17.26 & 14.64 & 13.92 & M5.75 \\ 
L366 &   2MASS J03443501+3208573 & 17.33 & 14.84 & 14.05 & M4.75 \\ 
L367 &   2MASS J03435915+3205567 & 17.36 & 14.68 & 13.95 & M5.75 \\ 
L373 &   2MASS J03442798+3205196 & 17.18 & 14.84 & 14.14 & M5.5   \\ 
L382 &   2MASS J03443095+3202441 & 18.95 & 15.48 & 14.47 & M5.5   \\ 
L391 &   2MASS J03444658+3209017 & 18.63 & 15.38 & 14.41 & M5.75 \\ 
L396 &   2MASS J03440233+3210154 & 17.57 & 14.98 & 14.18 & M5.25 \\ 
L405 &       -    		 & 18.34 & 15.20 & 14.48 & M8    \\ 
L414 &   2MASS J03444428+3210368 & 17.68 & 15.41 & 14.68 & M5.25 \\ 
L415 &   2MASS J03442997+3209394 & 18.43 & 15.20 & 14.36 & M6.5  \\ 
L432 &   2MASS J03444593+3203567 & 18.18 & 15.14 & 14.27 & M5.75 \\ 
L437 &   2MASS J03435638+3209591 & 18.61 & 15.41 & 14.62 & M7.25 \\ 
L454 &   2MASS J03444157+3210394 & 17.81 & 15.38 & 14.61 & M5.75 \\ 
L462 &   2MASS J03442445+3201437 & 19.18 & 15.67 & 14.58 & M3    \\ 
L468 &   2MASS J03441106+3201436 & 20.55 & 16.53 & 15.42 & M8.25 \\ 
L555 &   2MASS J03444121+3206271 & 16.86 & 14.28 & 13.54 & M5.75 \\ 
L603 &   2MASS J03443341+3210314 & 19.95 & 16.33 & 15.61 & M8.5  \\ 
L611 &   2MASS J03443035+3209446 & 19.61 & 16.35 & 15.49 & M8   \\ 
L613 &   2MASS J03442685+3209257 & 19.80 & 16.86 & 16.01 & M8.25 \\ 
L622 &       -    		 & 20.13 & 17.54 & 16.91 & M6    \\ 
L690 &   2MASS J03443638+3203054 & 20.02 & 16.62 & 15.78 & M8.75 \\ 
L703 &   2MASS J03443661+3203442 & 20.10 & 16.65 & 15.70 & M8    \\ 
L705 &       -    		 & 20.93 & 17.11 & 16.27 & M9   \\ 
L725 &       -    		 & 20.91 & 18.16 & 17.37 & M6   \\ 
L738 &       -    		 & 20.92 & 17.47 & 16.90 & M8.75 \\ 
L1683 &  2MASS J03441583+3159367 & - & - & - & M5.25 \\ 
L1684 &  2MASS J03442330+3201544 & 17.29 & 14.78 & 14.05 & M5.75 \\ 
L1889 &  2MASS J03442135+3159327 & - & - & - & -     \\ 
L1925 &  2MASS J03440576+3200010 & - & - & - & M5.5  \\ 
L4011 &      -    		 & - & - & - & -     \\ 
L4044 &      -    		 & 21.47 & 17.52 & 16.59 & M9    \\ 
L30003 & 2MASS J03435925+3202502 & - & - & - & M6    \\ 
\enddata
\tablecomments{\label{IC348objs}Identifications beginning with ``L'' are from the compilation of \citet{Luhman:2003p2777} and references therein,
as are the photometry and spectral types.}
\end{deluxetable*}

\begin{deluxetable}{cccccc} \tabletypesize{\scriptsize} \tablecolumns{6} \tablewidth{0pt} \tablecaption{IC~348 cluster members observed with the 
{\em Hubble Space Telescope.}} \tablehead{ \colhead{Object} & \colhead{$I$} & \colhead{$J$} & \colhead{$H$} &\colhead{SpT}
}                  
\startdata
L302  & 17.04 & 14.24 & 13.32 & M4.75 \\
L350  & 16.91 & 14.32 & 13.60 & M5.75 \\   
L405  & 18.34 & 15.20 & 14.48 & M8 \\
L761  & 20.03 & 15.66 & 15.33 & M7 \\
L1434 & 21.11 & 18.39 & 17.44 & M6 \\
L4044 & 21.47 & 17.52 & 16.59 & M9 \\
\enddata
\tablecomments{\label{IC348HSTobjs}All information is from the compilation of \citet{Luhman:2003p2777} and references therein.}
\end{deluxetable}

\subsection{Upper Scorpius}

The Upper Scorpius (USco) region is one of the most spatially extended young associations, with stars 
and BDs spread over many tens of degrees on the sky. As a result, few variability studies have been 
performed here, apart from the work of \citet{2008PhDT.......439S}. It is significantly older than the
other three regions studied, with age estimates from 5 to 11~Myr \citep{Preibisch:2002p4082, 2012ApJ...746..154P}. 
Nevertheless, an examination of the H-R diagram of catalogued low-mass members shows that
many objects have temperatures and luminosities overlapping the D-burning instability strip. 

Because of the sparseness of this association, it is difficult to obtain data on more than one target at a time. 
We therefore selected fields carefully to maximize the number of pulsation candidates.
We ultimately observed five different fields in Upper Scorpius, including 5 BDs
and 11 VLMSs, with identifications and references listed in 
Table~5. None of these regions contained any nebulosity.
Observations of three of the fields were abbreviated to three nights or less 
because of weather (the CTIO~1.0~m run), and tracking problems (USco members in the first field chosen for 
observation with the P60 fell too close to the edge of the detector
and tended to wander out of the FOV). The list of dates is provided in Table~1.
Since our observations in 2008 and 2009, further low-mass Upper Scorpius members have been discovered 
by \citet{Lodieu:2011p7855} and \citet{Dawson:2011p4584}. We identified three of these objects from 
\citet{Lodieu:2011p7855} in our first FOV from CTIO~1.0~m Y4KCam
monitoring in May 2008, with no additional targets in any of the other observations. 

\begin{deluxetable*}{cccccc}
\tabletypesize{\scriptsize}
\tablecolumns{6}
\tablewidth{0pt}
\tablecaption{Objects in Upper Scorpius observed as part of the pulsation campaign.}
\tablehead{
\colhead{Object} & \colhead{Telescope} & \colhead{$i$} & \colhead{$J$} & \colhead{SpT} &
\colhead{Reference} 
}
\startdata
DENIS-P-J161050.0-221251.6 & CTIO~1.0~m  & - & 12.80  & M5.5 & 1 \\
UScoJ16111705-2213088 & CTIO~1.0~m  & - & 11.64 & M5  & 2 \\
SCH~J16111711-22171749 & CTIO~1.0~m  & 17.97  & 14.34  & M7.5  & 3 \\
UScoJ16113470-2219443 & CTIO~1.0~m  & - & 13.24 & M5.75 & 2 \\
UScoJ16113784-2210275 & CTIO~1.0~m  & - & 11.07 & M4  & 2 \\
SCH~J16115737-22150691  & CTIO~1.0~m  & 16.70  & 13.73  & M5 & 3 \\
SCH~J16130306-19293234 & P60      & 16.75  & 13.45  & M5.5 & 3 \\
SCH~J16132809-19245288 & P60      & 16.16  & 12.92  & M6  & 3 \\
SCH~J16172504-23503799  & CTIO~1.0~m  & 17.20 & 13.74 & M5   & 3 \\
SCH~J16173105-20504715  & P60     & 16.49  & 13.03 & M7   & 3 \\
SCH~J16174540-23533618 & CTIO~1.0~m  & 17.44 & 14.05 & M6   & 3 \\
SCH~J16181567-23470847 & CTIO~1.0~m  & 16.18 & 12.42 & M5.5 & 3 \\
SCH~J16182501-23381068 & CTIO~1.0~m  & 17.19 & 13.72 & M5   & 3 \\
SCH~J16183144-24195229 & P60      & 17.76 & 14.15 & M6.5 & 3 \\
SCH~J16183620-24253332 & P60      & 14.75 & 12.03 & M4 & 3 \\
SCH~J16185038-24243205 & P60      & 16.79 & 13.63 & M5 & 3 \\
\enddata\tablecomments{\label{UScoobjs}References: (1) \citet{Martin:2004p4210}; (2) \citet{Lodieu:2011p7855}; (3) \citet{Slesnick:2008p4044}}
\end{deluxetable*}

\section{Data acquisition and reduction}

We employed four different telescopes in the search for pulsation, as summarized in Table~1. An extensive discussion of 
observations with the {\em Spitzer} Space Telescope is provided in Cody \& Hillenbrand (2011); here we detail the observing
set-ups with the other three facilities.

\subsection{Cerro Tololo Interamerican Observatory 1.0m telescope}

We observed low-mass targets in $\sigma$~Orionis, Chamaeleon~I, and
Upper Scorpius with the CTIO~1.0m.  The observational set-up and data reduction procedures were the nearly same
for all CTIO~1.0m runs, and they are described in detail in CH10. The selected exposure time was 600 seconds for all
runs, apart from the first set of observations on $\sigma$~Orionis
members (360 seconds). 

The main distinction for the May 2008 observing run (see Table~1) was that sky conditions were not photometric, and just over two nights 
were lost to clouds. In addition, telescope building maintenance caused a new accumulation of dust specks 
on the detector each night, which resulted in inconsistent sky flatfield acquisition. We acquired dome flatfields at 
the beginning and end of each night to calibrate out dust ``donuts'',
but these dome flats are known to misrepresent 
the true pixel sensitivity distribution by up to 10\%. Therefore, we carried out 
flatfielding with sky flats on nights where at least seven were available, and when clouds precluded the acquisition of 
sky flats, we instead relied on the dome flats but performed an illumination correction using the high 
signal-to-noise composite provided by 
P.~Massey\footnote[2]{http://www.lowell.edu/users/massey/obins/y4kcamred.html}.

\subsection{Palomar 60-inch telescope}
\label{P60reduction}

We used the P60 robotic telescope to
observe BDs and VLMSs in the IC~348 cluster, as well as additional BDs
in Upper Scorpius. Observations of IC~348 took place on a total of 9 nights between 2008 November 17 and 
November 29. The chosen field center was ${\rm R.A.}=03^{\rm h}$44$^{\rm m}$19.7$^{\rm s}$, 
${\rm decl.}=+32\arcdeg$04$\arcmin$29$\arcsec$s (J2000), but since the
P60 system was at that time subject to tracking inaccuracies, this position shifted
up to 45$\arcsec$ throughout the run. We observed USco with the
P60 during two different runs, from 2008  June 1--14, as well as 2009 May 14--30.

We chose exposure times of 240 or 300 seconds in the $i'$ band to provide good sensitivity to the faint brown dwarfs without elongating the PSFs too much 
due to the lack of guiding. The established data reduction pipeline
for the P60 performs basic calibrations, including bias 
subtraction, flatfielding, and fitting of the world coordinate system \citep{Cenko:2006p5614}. Although we 
obtained a series of sky flatfields, we determined that the domeflat images used by the pipeline were 
sufficient to correct interpixel sensitivity variations. Image alignment was also carried out with ease, since an accurate coordinate system was already 
superimposed on the calibrated images; we used the {\em IRAF} task {\em wregister} to complete this 
task.

\subsection{HST}

We used the {\em HST} WFC3 ultraviolet/visible (UVIS) CCD to observe
pulsation candidates in IC~348. The UVIS channel is comprised of two chips, each with 4096~$\times$~2051 pixels; 
since we observed in subarray mode, we used only one of these (UVIS1). Each pixel is $\sim 0.04\arcsec$ across, 
for a total subarray field of view of $\sim$81$\arcsec$~$\times$~162$\arcsec$. 

Observations took place from 2011 January 29 to February 4, for just over 7 hours of each day. Although {\em 
HST} is a space observatory, the sun position and other observing constraints resulted in each block of 
observations (``visit'')  beginning at roughly the same time every day. Unfortunately much of visit 5 was 
compromised since the gyroscopic system failed and the field was lost for a number of hours. The viewing 
limits of {\em HST} are such that IC~348 objects may be observed for only 46 minutes of each 97 minute 
orbit. Therefore we designed exposure times to acquire as many images as possible per orbit, without 
exceeding the telescope's maximum data downlink rate. These varied among 128, 171, and 192 seconds.  All 
observations were carried out through the F814W filter, which is centered near 8030~\AA\ and similar to $I$ 
band.

{\em HST}/WFC3 data are processed by pipeline, which includes standard bias and flatfield calibration, as 
well as cosmic ray rejection. The MultiDrizzle program corrects for geometric distortion and optimally 
combines sets of three of four consecutive images, even for undithered data such as ours.

\section{Photometry routines}
We performed aperture photometry on our target objects using 
a variety of aperture sizes and sky annulus widths and radii for background subtraction.
Different approaches were chosen to produce ground and space-based photometry. For 
the P60 and CTIO~1.0m data, we employed a variable-aperture method to optimize flux measurements in a variety of seeing 
conditions; see Cody \& Hillenbrand (2010) for a description of this
technique. Both standard aperture photometry and image subtraction
algorithms were tested. Typical apertures were 1--2 times the full width at half
maximum (FWHM) PSF size, depending on object brightness, and sky annulus radii from 4.5--6
times the FWHM were used to subtract background. To locate photometric
non-variable reference stars, we identified suitable field objects. Typically, we found at least
three to four references with photometric stability of several
millimagnitudes over the full duration of each run. 

For the {\em HST}/WFC3 images, apertures were fixed for each target and several different sizes were tested, 
from 2 to 8 pixels, as were sky annulus radii from 8--12 pixels and 11--16 pixels. Although results did not 
differ much, the best RMS light curve values were attained with the 6 pixel aperture and the sky annulus 
extending from 11--16 pixels. Although the Poisson errors are close to 0.001 magnitudes for several targets, 
the measured RMS light curve spreads are an order of magnitude larger, regardless of the type of photometry 
employed. This discrepancy suggests that the BDs and VLMSs observed
with {\em HST} are all intrinsically variable. 

The light curves produced for all fields were exclusively differential, since 
our primary interest in short-term flux variations makes photometric zero points irrelevant. 
By eye, the resulting time series from all telescopes display copious
periodic and erratic variability, the latter of which is likely associated with accretion and circumstellar processes (see CH10 and
the Appendix). 
To evaluate the performance of our photometric routines, we omitted
variables with RMS values of more than three standard deviations above
the median trend as a function of magnitude. Careful attention to ground-based
reduction procedures enabled us to reach a floor of several millimagmitude photometry
at the bright end, comparable to the performance from the space observatories.
In general, signal-to-noise, and hence photometric performance, decreased into the brown dwarf regime. 
We provide details on the data quality for individual clusters below in Section~5.

The final light curves were cleansed of artificial variability caused
by, e.g., crossings of bad pixel columns and detector malfunctions. In the case
of Cha~I, a number of reference stars were unavailable on the first
night of the run, since the two bottom quadrants of the Y4KCam CCD had
failed. Not counting
the first night, most Cha~I light curves had 278 
datapoints for the entire run. To recover additional photometry on the
first night for stars in the top of the FOV,  we performed image
subtraction, which does not rely on reference stars. This resulted in
an additional 26 datapoints for about half the targets. Light curves
for the other clusters contained between 31 and 528 points. 

\section{Periodic variability characterization}

\begin{figure*}
\begin{center} 
\includegraphics[scale=0.8]{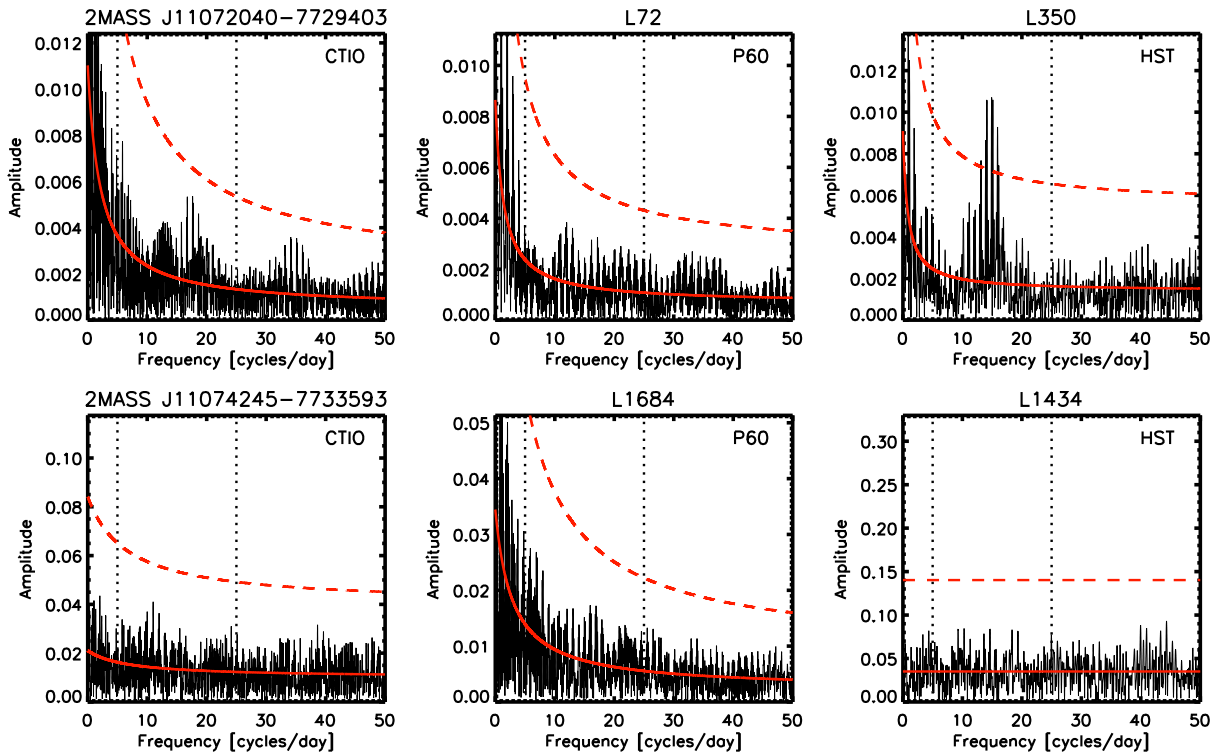}
\end{center}
\caption[]{\label{periodos}Example periodograms for bright (top row)
 and faint objects (bottom row), for data acquired from each
 telescope involved in the pulsation search. Vertical dashed lines
indicated the region of frequency space ($\sim$5--25 cd$^{-1}$) where pulsation is
predicted. Solid red curves mark $1/f$ profile fits to the noise level. The
upward trends toward low frequency are generally due to intrinsic
variability on $>$1 day timescales. Dashed red curves are four times
the noise level, which is our adopted limit for detection of
significant signals. The prominent set of peaks between 10 and 20
cd$^{-1}$ in the periodogram of L350 are aliases due to the 97 minute
orbital period of {\em HST}.}
\end{figure*}

Our main aim in analyzing the light curves of low-mass cluster members was to
search for periodicities on the 1--4-hour timescales predicted for deuterium-burning
instability. We produced Fourier transform periodograms
\citep{1975Ap&SS..36..137D}, which display the amplitude of
periodic signals as a function of frequency in units of cycles per day
(cd$^{-1}$), along with the spectrum of noise in the light
curve. Examples are displayed in Figure~\ref{periodos}. We find that most display an
exponential rise toward low frequencies, indicative of systematic
effects (both night-to-night photometry shifts and intrinsic aperiodic
stellar variability) on $>$1 day timescales in the light curve.
Each periodogram is computed from a lower bound of $1/T$, where $T$ is
the total duration of observations, up to the Nyquist limit, or one half
of the sampling rate (typically $\sim$70 cd$^{-1}$).

True periodicities appear as localized peaks in the periodogram, and
we identify them as statistically significant if they rise higher than
a factor of 4.0 times the surrounding local noise level (see CH10 for
further discussion on this requirement). We determine the noise profile by
fitting a curve of form $A/(f+B)+C$ curve to the periodogram values as
a function of frequency, $f$, with constants $A$, $B$, and $C$. This
is displayed as the red curves in Figure~\ref{periodos} and confirms that the minimum 
detectable variability level at low frequencies is somewhat larger than amplitudes observable at higher frequencies 
(i.e., shorter timescales).

We searched for periodicities
in the periodograms corresponding to each light curve, paying
particular attention to the 5--25~cd$^{-1}$ range (i.e., periods of
1--4.8~h) predicted for deuterium burning pulsation. If a period was
found, we then removed the best-fit trend \citep[based on multi-sine fits with the program Period04;][]{Lenz:2005p7898} to 
produce a pre-whitened light curve. The search for few-hour periodicities was then carried out on this residual. 

All targets were subjected to periodogram analysis. As a separate exercise, aperiodic variables were also
identified; these ``contaminants'' are discussed in the Appendix.
In the following sections, we describe the results of our period
searches in each of the $\sigma$~Orionis, Cha~I, IC~348, and Upper Scorpius regions. 

\subsection{$\sigma$ Orionis}

With two nearly two-week observing campaigns on the CTIO~1.0m
telescope, we monitored over 150 VLMSs and BDs in the $\sigma$~Orionis
cluster. As detailed in CH10, we identified 38 periodically variable confirmed cluster
members, along with 27 periodically variable candidate
members. However, all but one of these variables had periods greater
than 8 hours. The remaining single object (2MASS J05382557−0248370) had a period of 7.2 hours.
This lack of detections in the 1--4 hour range was despite sensitivity to periods as short as 15 minutes, and
amplitudes as low as several millimagnitudes down to a magnitude of
$I$=17 (and as low as 1 mmag at $I$=14). We therefore concluded that
there are no signs of deuterium-burning pulsation in our $\sigma$~Ori dataset.
The full compilation of periodic variables, their phased light curves,
and discussion of detection limits appears in CH10.

\subsection{Chamaeleon I}

Using data from the CTIO~1.0m telescope, we produced and analyzed light curves for all 32 observed BD and 
VLMS members of the Cha~I cluster. In the interest of fully mining the dataset and potentially 
identifying new cluster members, we additionally performed photometry on all 1548 objects in the Cha~I 
field that were bright enough for detection in individual Y4KCam images ($i\lesssim 22$).

\begin{figure}
\begin{center}
\includegraphics[scale=0.5]{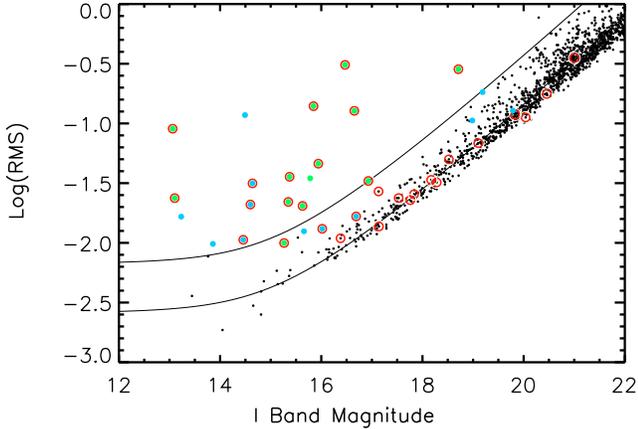}
\end{center}
\caption[]{\label{apermagrms}RMS spread of light curves over the
13-night duration of the Cha~I observations, for periodic (blue) and aperiodic (green) variables.
Confirmed cluster members appear as red circles. We plot the estimated total contributions from
Poisson, mean sky level, and systematic noise, shifted upward by 0.24 dex so as to match the median of the
data (lower solid line). The curve corresponding to 99\% probability of variability detection via the $\chi^2$ 
test appears above this.
}
\end{figure}

These additional light curves enabled an assessment of the
photometric noise as a function of magnitude, as shown in
Figure~\ref{apermagrms}. We have fit a trend to the cluster
non-members, based on the expectations for Poisson and sky
noise. Using the fit values as an indication of the magnitude-dependent photometric
uncertainty, we have then selected
variable objects via the $\chi^2$ test. The full collection of these light curves
is presented and discussed in the Appendix.
The floor of the RMS distribution reaches $\sim$3~mmag at an $I$-band magnitude of
14, as shown in Figure~\ref{apermagrms}. Objects with $I<17$ display a photometric precision of 1\% or better.

For each object in the sample, we carefully analyzed the periodograms for signs of periodicities on the 
few-hour timescales predicted for D-burning pulsation. Among the
known cluster members, only four stars have estimated masses above 
$\sim$0.4~$M_\odot$, and so most are candidates for the instability. 

Based on the observing cadence of 600s, we are sensitive to periodic signals with frequencies as high as 
72~cd$^{-1}$ (i.e., $P$=20 minutes). The majority of light curves contain low-frequency variations, most 
of which are due to intrinsic erratic variability; these shape the exponential rise in periodogram 
amplitude at $<$1~cd$^{-1}$. Despite sensitivity to few-millimagnitude levels, we do not find any 
evidence for variability with periods less than 16 hours, apart from one field object that appears to be 
a main sequence pulsator\footnote[3]{The star 2MASS J11105665-7733557 has a period of 3.3 hours and 
amplitude 0.12 magnitudes in the $I$ band. However, its near-infrared colors lie well blueward of the 
young cluster sequence, implying that it is not a Cha~I member}. A few periodograms display low-level 
signals in the range where pulsation is expected, but none of these meet the 99\% significance level 
criteria, and we find most to be aperiodic variables. Furthermore, the light curves do not show clean 
trends when phased to these periods. We conclude that none of the 32 very low mass objects in the Cha~I 
sample are periodic on 1--5 hour timescales, at least at or above the amplitude levels probed by the 
data.

\begin{figure*}
\begin{center} 
\includegraphics[scale=0.8]{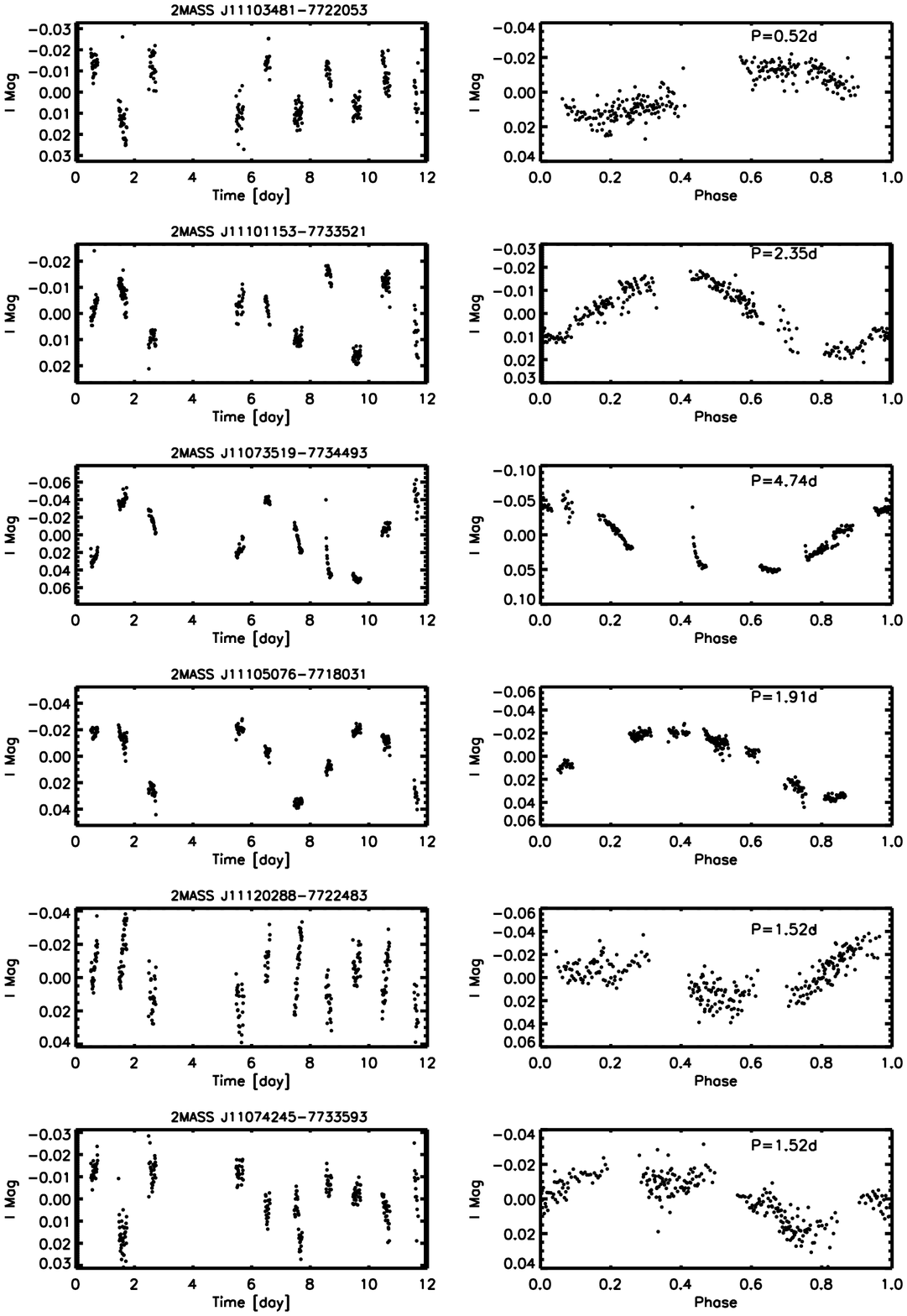}
\end{center}
\caption[]{\label{chaIperlightcurves}Differential light curves for very-low-mass
 Cha~I members with detected periodic variability. 
The first column shows the original light curve, while the second is
the light curve phased to the detected period.}
\end{figure*}

\begin{figure*}
\begin{center}
\includegraphics[scale=0.8]{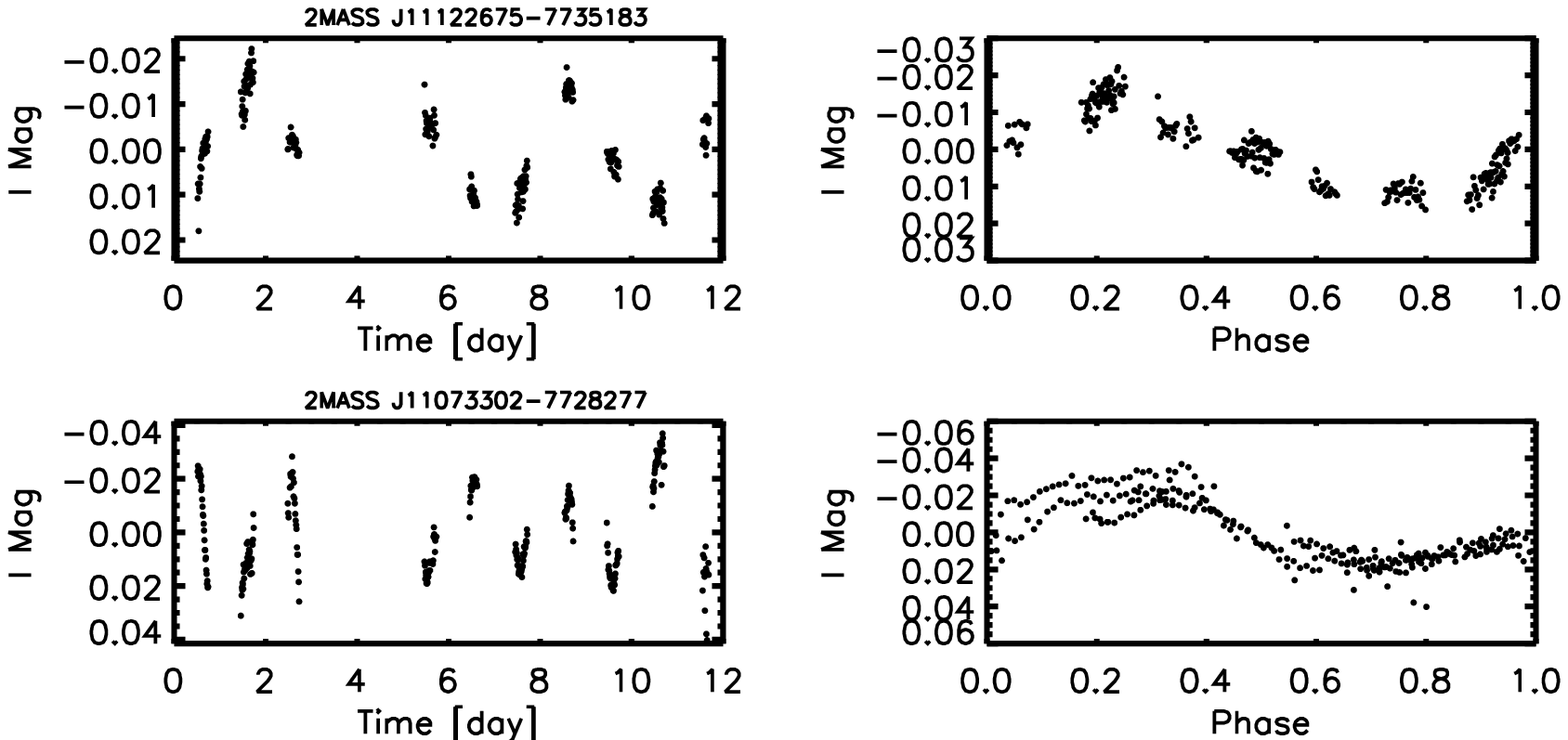}
\end{center}
\caption[]{\label{chaIcandperlightcurves}Light curves for
new candidate Cha~I members with detected periodic variability. 
The first column shows the original light curve, while the second is
the light curve phased to the detected period.}
\end{figure*}

On longer timescales, the final sample of variables contains 12 periodic objects in Cha~I with clear 
variability. Observed and phased light curves for the cluster members among these are 
presented in Figure~\ref{chaIperlightcurves}. We also identified periodic behavior among eight additional 
objects of unknown membership status. In most of these cases, the light curve shapes are characteristic 
of field pulsators or eclipsing binaries, and blue colors suggest locations in the background field. 
However, two relatively red objects with sinusoidal light curves may be bona fide Cha~I 
members (2MASS J11073302-7728277 and 2MASS J11122675-7735183). The former (also known as CHXR 25) was 
classified by \citet{Luhman:2007p2372} as a field dwarf, while the latter has no previous literature
and would benefit from spectroscopic follow-up. The light curves for these new Cha~I candidates are presented 
in Figure~\ref{chaIcandperlightcurves}.

The measured properties for all variable sources in Cha~I are listed in Table~6. This includes the aperiodic variables
uncovered with the $\chi^2$ test, as discussed in the Appendix.
Of note, few Cha~I members have been photometrically monitored previously, apart from a sample of 10 BDs and VLMSs 
presented by \citet{Joergens:2003p3530}. Of the five with periods reported in that work, {\em none} are redetected as 
periodic variables here.

\begin{deluxetable*}{ccccccc}
\tabletypesize{\scriptsize}
\tablecolumns{5}
\tablewidth{0pt}
\tablecaption{Cha~I objects with detected variability}
\tablehead{
\colhead{Object} & \colhead{Variability type}  & \colhead{Period} &
\colhead{Amplitude} & \colhead{RMS} &
\colhead{Disk?} & \colhead{Member?} 
}
\startdata
2MASS J11065906-7718535 & A & - &0.405 & 0.090 & Y & Y\\
2MASS J11071668-7735532 & A & - &0.405 & 0.090 & Y & Y\\
2MASS J11072040-7729403 & A & - &0.105 & 0.022 & N & Y\\
2MASS J11072988-7725017& P & 2.28d & 0.25 & 0.117 & - & N\\
2MASS J11073302-7728277 & P & 0.67d & 0.0197 & 0.017 & N  & M\\
2MASS J11073519-7734493 & P & 4.74d  & 0.0478 & 0.031 & N & Y\\
2MASS J11073686-7733335 & A & & 0.098 &0.020 & N & Y\\
2MASS J11074245-7733593 & P & 1.52d & 0.0138 & 0.012 & -  & N\\
2MASS J11075225-7736569 & A & - & 0.090 & 0.022 & N & Y\\
2MASS J11082238-7730277 & A & - & 0.229 & 0.046 & Y & Y\\
2MASS J11083952-7734166 & A & - & 0.145 & 0.036 & Y & Y\\
2MASS J11085421-7732115 & A & - & 0.043 & 0.010 & N & Y\\
 CTIO J11093360-7731113 & P & 0.46d & 0.1271 & 0.128 & - & N\\
2MASS J11094742-7726290 & A & - & 0.571 & 0.128 & Y & Y\\
2MASS J11094918-7731197 & A & - & 0.128 & 0.026 & N & Y\\
2MASS J11101153-7733521 & P & 2.354d & 0.0152 & 0.011 & N & Y\\
2MASS J11103481-7722053 & P & 0.5229 & 0.0137 & 0.013 & N & Y\\
2MASS J11105076-7718031 & P & 1.91d & 0.0286 & 0.021 & N & Y\\
2MASS J11105359-7725004 & A & - & 1.431 & 0.285 & Y & Y\\
2MASS J11105665-7733557 & P & 0.14d & 0.1244 & 0.106 & -  & N\\
 CTIO J11111463-7737020 & P & 0.59d & 0.1827 & 0.183 & - & N\\
2MASS J11120288-7722483 & P & 1.52d & 0.0169 & 0.017 & N & Y\\
2MASS J11120351-7726009 & A & - & 1.122 & 0.309 & Y & Y\\
2MASS J11122675-7735183 & P & 3.52d & 0.0129 & 0.010 & N & M\\
2MASS J11122971-7731045 & A & - & 0.171 & 0.035 & - & M\\
\enddata
\tablecomments{\label{chaIvartable}We list objects in the Cha~I field with detected variability. ``A'' 
corresponds to aperiodic variability, while ``P'' is for periodic variability. The values listed in
column 4 are either the half amplitude for periodic sources or the
peak-to-peak amplitude for aperiodic sources. In column 5 we provide 
the RMS light curve spreads in $I$-band magnitudes. The disk
column indicates whether mid-infrared {\em Spitzer} data exhibits an
excess; ``-'' indicates a lack of {\em Spitzer} photometry. Membership
is based on previous censuses of Cha~I; non-members (``N'') have
colors that are inconsistent with a position above the main sequence. 
``M'' indicates a possible new cluster member based on the detected variability.}
\end{deluxetable*}

\subsection{IC 348}

\begin{figure}
\begin{center}
\includegraphics[scale=0.5]{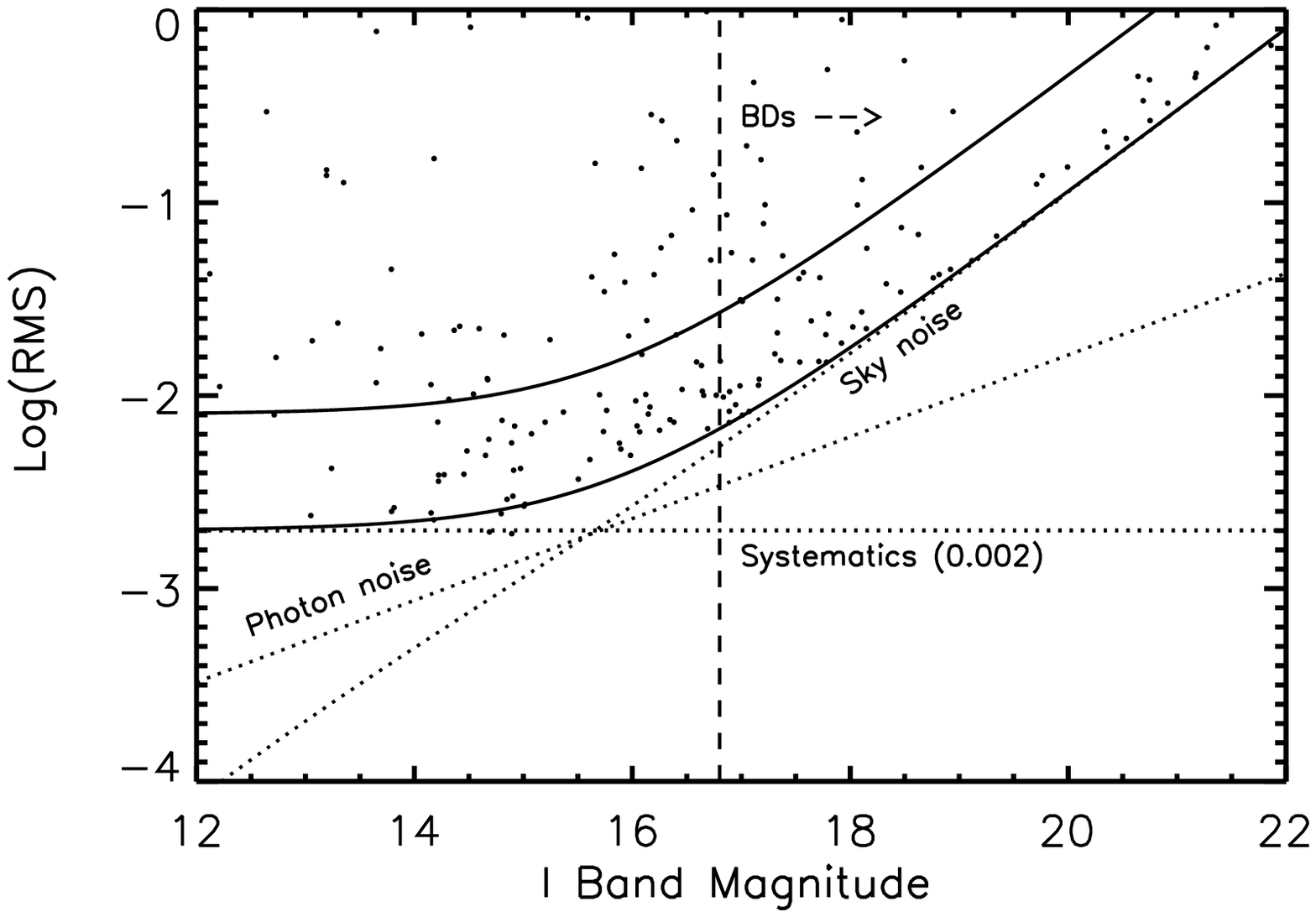}
\end{center}   
\caption[]{\label{apermagrms1}RMS spread of light curves for sources in
IC~348, as measured over the course of a single night. All plotted
objects are confirmed cluster members. We show the estimated total
contributions (solid line) from Poisson, mean sky level, and
systematic noise. The upper solid curve corresponds to an estimate for
99\% probability of variability detection. There is no clear boundary 
between variable and non-variable sources here, unlike in Figure~\ref{apermagrms},
which contains a mix of young cluster members and field stars.}
\end{figure}

In IC~348 the photometry includes only known cluster members, since high extinction in
the region blocks our view of most background field stars. We were
therefore unable to perform a full assessment of photometric
precision based on non-variable field stars. However, a plot of RMS light curves values versus magnitude
(shown in Figure~\ref{apermagrms1}) provides a rough estimate of the
performance for our P60 dataset. Over the course of a single night, we reach a precision
of $\sim$2 millimagnitudes at $I$=13, and better than 0.01
magnitudes down to $I\sim17$. The RMS values measured over the entire
week-long duration of the light curves show considerably higher
spreads in photometry, $\sim1$\% or higher in most light curves. We
attribute much of this to intrinsic variation of the stars on $>$1 day timescales.

We performed a period search analysis on all IC~348 cluster members 
observed with the P60 and {\em HST}. For the former, we are sensitive to periodicities on timescales 
from approximately 9 minutes to 12 days (resulting from a cadence of 270s). Since we did not obtain data every night, our sensitivity to 
periods of more than a few days is highly non-uniform. We therefore focused on the short periods predicted for 
D-burning pulsation. 

We generated periodograms from the light curves of all 147 unsaturated objects in the 
ground-based P60 field, including 91 BDs and VLMSs. A large fraction of these display variability by eye, much of which is erratic and 
dominated by low-frequency behavior. It is thus not surprising that many of the periodograms 
display excess power around 5--10~cd$^{-1}$, which we attribute to
aliasing, based on the window function, which shows a series of peaks
centered around 8~cd$^{-1}$. Further support for the spurious nature of these signals is that they
do not reach the 99\% significance level, and no coherent periodicity
emerges when the light curve is phased. 

In some cases, an obvious and statistically significant periodicity
did appear on timescales of one day or more, suggesting rotation. When this occurred, we fit the overall trend and removed 
it from the light curve before searching for shorter timescale pulsation signals. 
In {\em none} of the IC~348 light curves did we uncover significant
periodicities on the 1--5 hour timescales indicative of pulsation.

Similar to the ground-based monitoring data, the {\em HST} time series
are subject to aliasing. This effect is associated with the 97-minute
orbital timescales and is seen prominently in the periodogram of
object L350, shown in Figure~\ref{periodos}. Omitting the
frequency regions subject to aliasing, we have again searched for periodicities on 1--5 hour timescales, in hopes of 
detecting pulsation.  In general, we find that the periodogram values
are consistent with noise at the 1--5 millimagnitude level in the
frequency range of interest (5--25~cd$^{-1}$). 

\begin{figure}
\begin{center}
\includegraphics[scale=0.5]{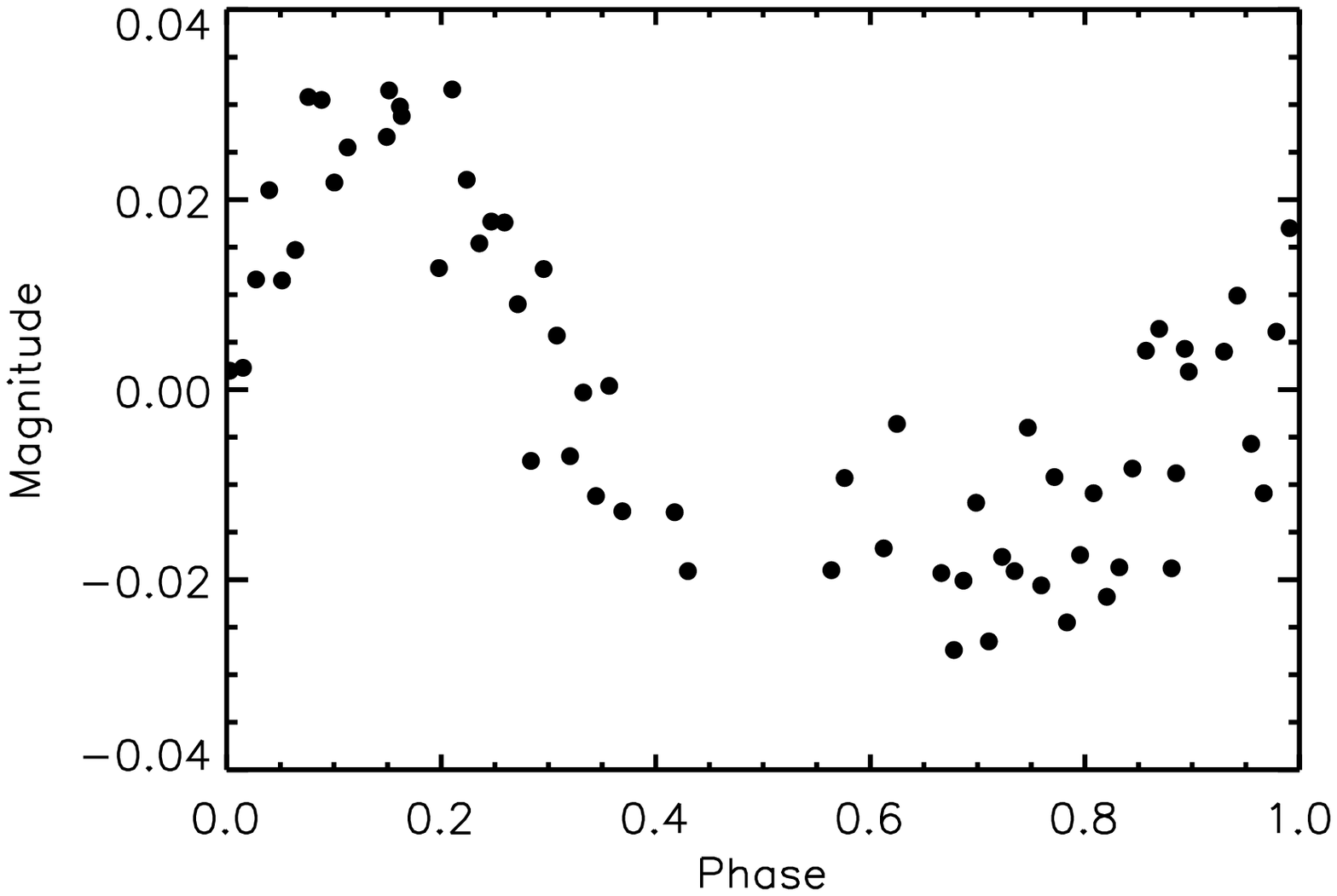}
\end{center}
\caption[]{\label{l761}Phased light curve of the IC~348 object L761, observed with {\em HST}/WFC3. This BD displays
significant periodic variability on a timescale of 0.6 days.}
\end{figure}

\begin{figure}
\begin{center}
\includegraphics[scale=0.5]{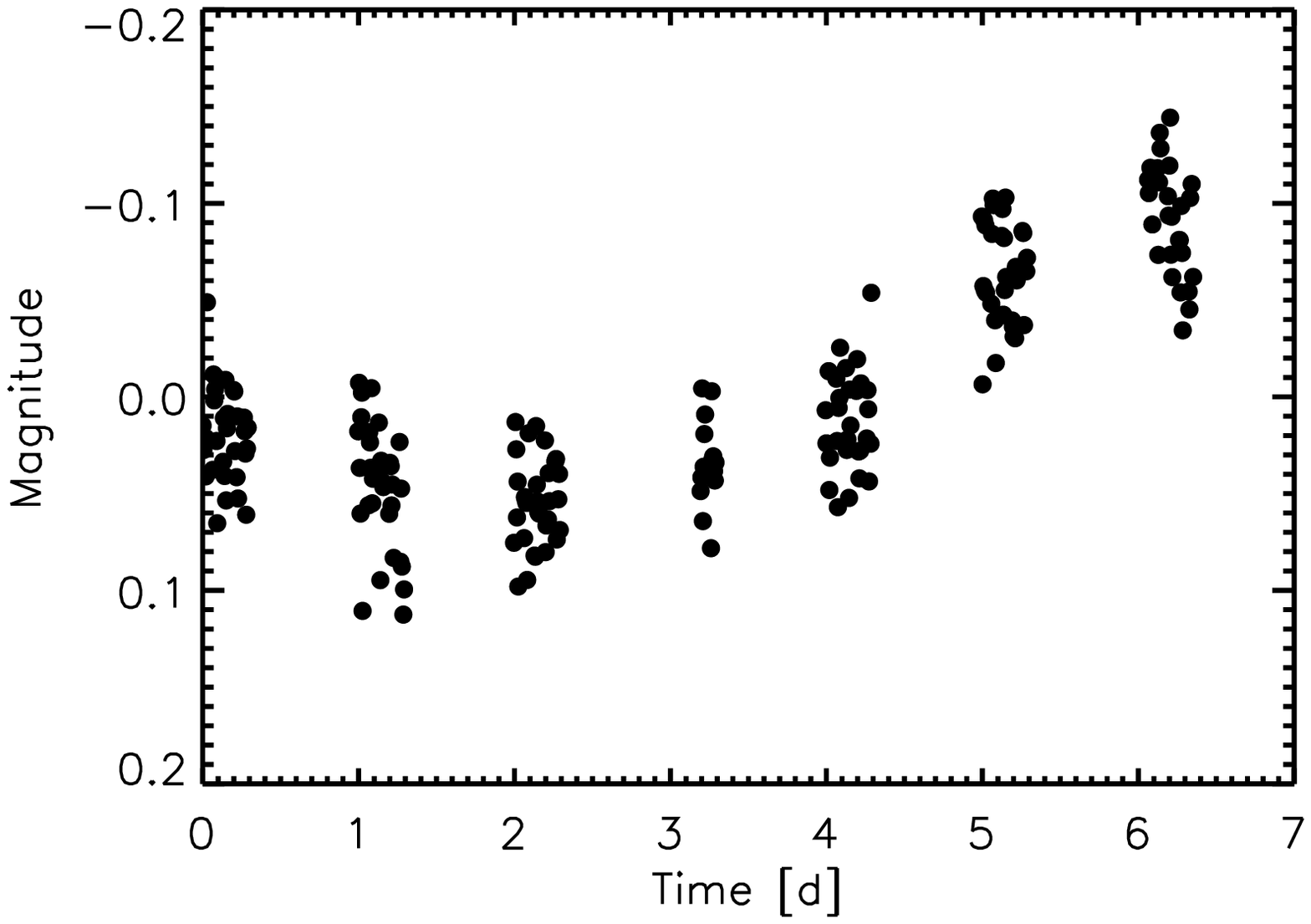}
\end{center}
\caption[]{\label{l1434}Light curve of the IC~348 object L1434, observed with {\em HST}/WFC3. This BD displays 
significant variability over the 7 days that it was monitored.}
\end{figure}

Two exceptions in the {\em HST} dataset are the IC~348 brown dwarfs
L761 and L1434. The former is periodically variable, as shown in Fig.~\ref{l761}. 
The period of 0.6 days is consistent with rotational
modulation by one or more spots on its surface. 
L1434 is also variable (see Figure~\ref{l1434}),
but we are unable to determine whether it is periodic without a longer
duration dataset.

\vspace{0.5cm}
\subsection{Upper Scorpius}

As with the other young star regions, we produced discrete Fourier transforms to search for periodicities on
a variety of timescales in Upper Scorpius. Targets here were observed
during a number of runs (see Table~1), and we are sensitive
to periodicities as short as 10 minutes in each of these datasets. The longest 
period detectable varies with the length of each observing run and ranges from less than two days to 16 days. Since 
some of the runs only had six or fewer total nights, the associated periodograms have lower frequency 
resolution than the datasets on other clusters, and the search for periodicities is more susceptible to 
systematic effects. While some of the objects display variability on night-to-night 
timescales, we cannot accurately quantify all possible
periodicities. We therefore focused exclusively on the search for pulsation at 1--5 
hour periods. Detection of signals in this period range is feasible given the number of datapoints (30--550, depending on the run) 
and the fairly low photometric uncertainties of the brown dwarfs (1--3\%).

We find that the data generally have sensitivity to
signals in the few millimagnitude to 0.01 magnitude amplitude range.
Despite this, none of the objects in USco showed significant
periodic variability. However, the light curves of a few displayed
night-to-night variations that may be indicative of accretion or
variable circumstellar extinction. 

\section{Non-detection of the deuterium burning instability}

With no periodicities observed in the 1--4 hour range for
any of the clusters monitored, we now consider the implications of our non-detections. 
In each region, we assess the set of
temperatures and luminosities in relation to the deuterium burning
instability strip on the H-R diagram. The number of objects on or near the
theoretically predicted region then sets the probability of detection, or lack thereof. 

\begin{figure*}
\begin{center}
\epsscale{1.0}
\plottwo{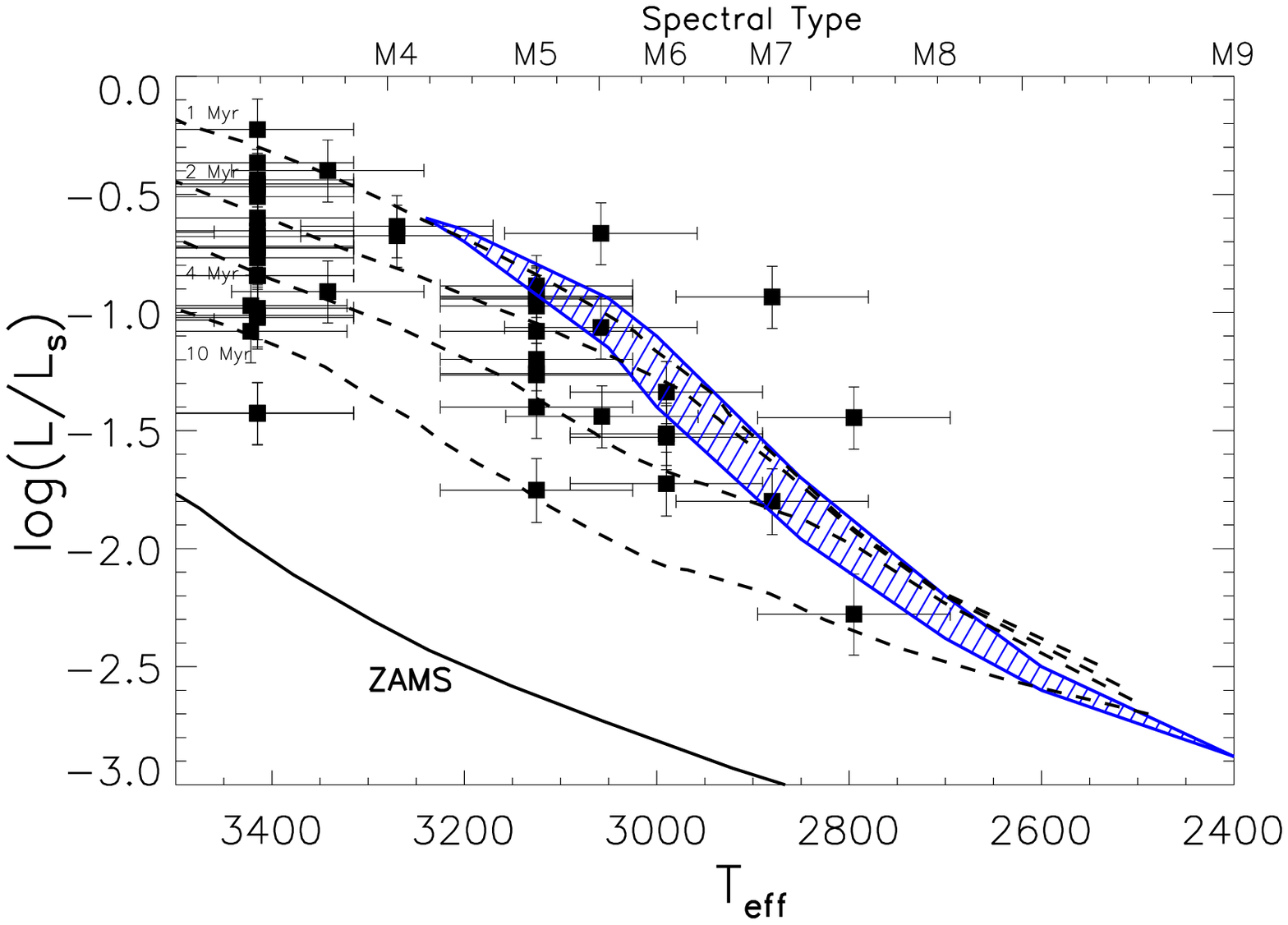}{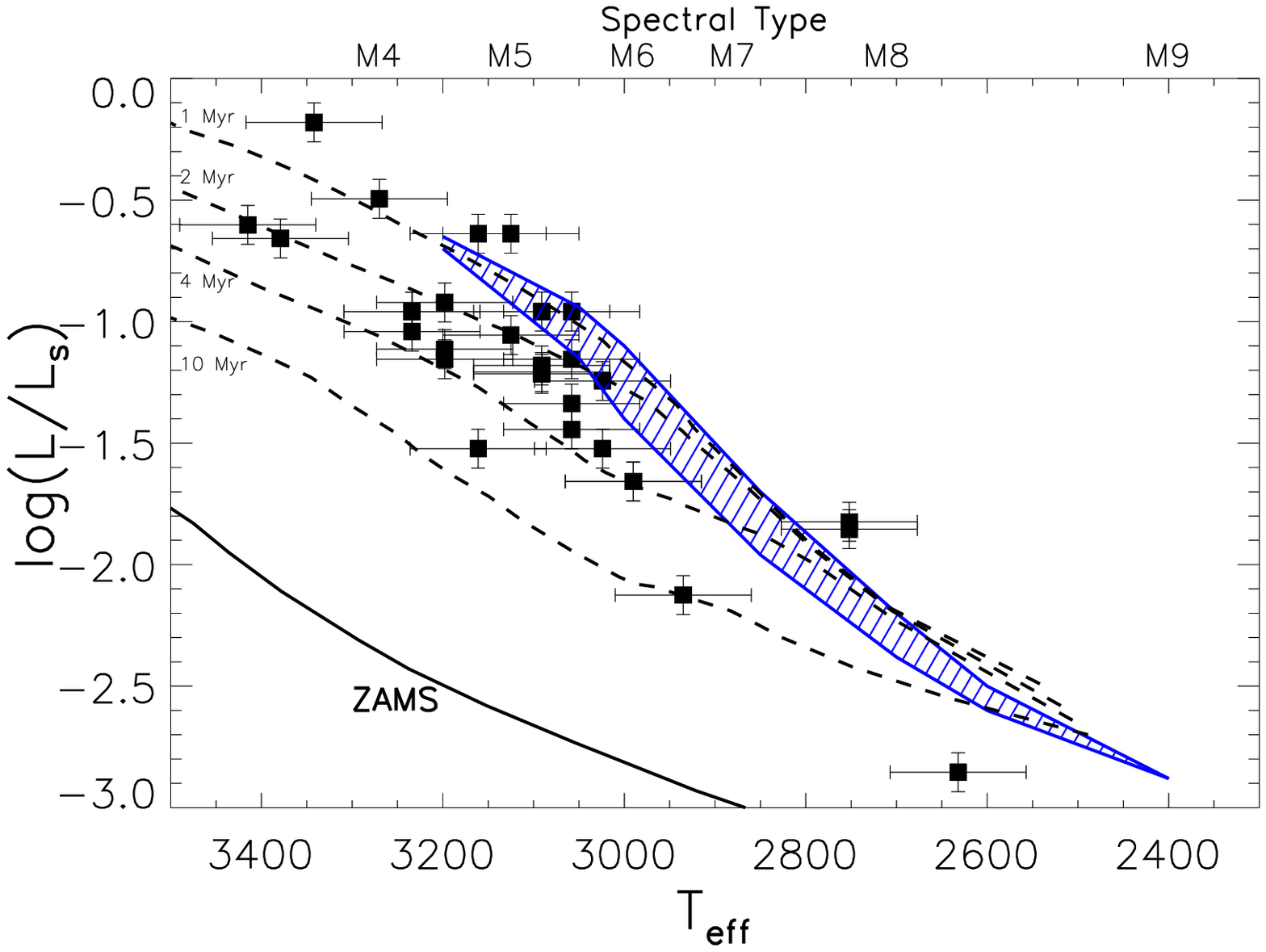}
\plottwo{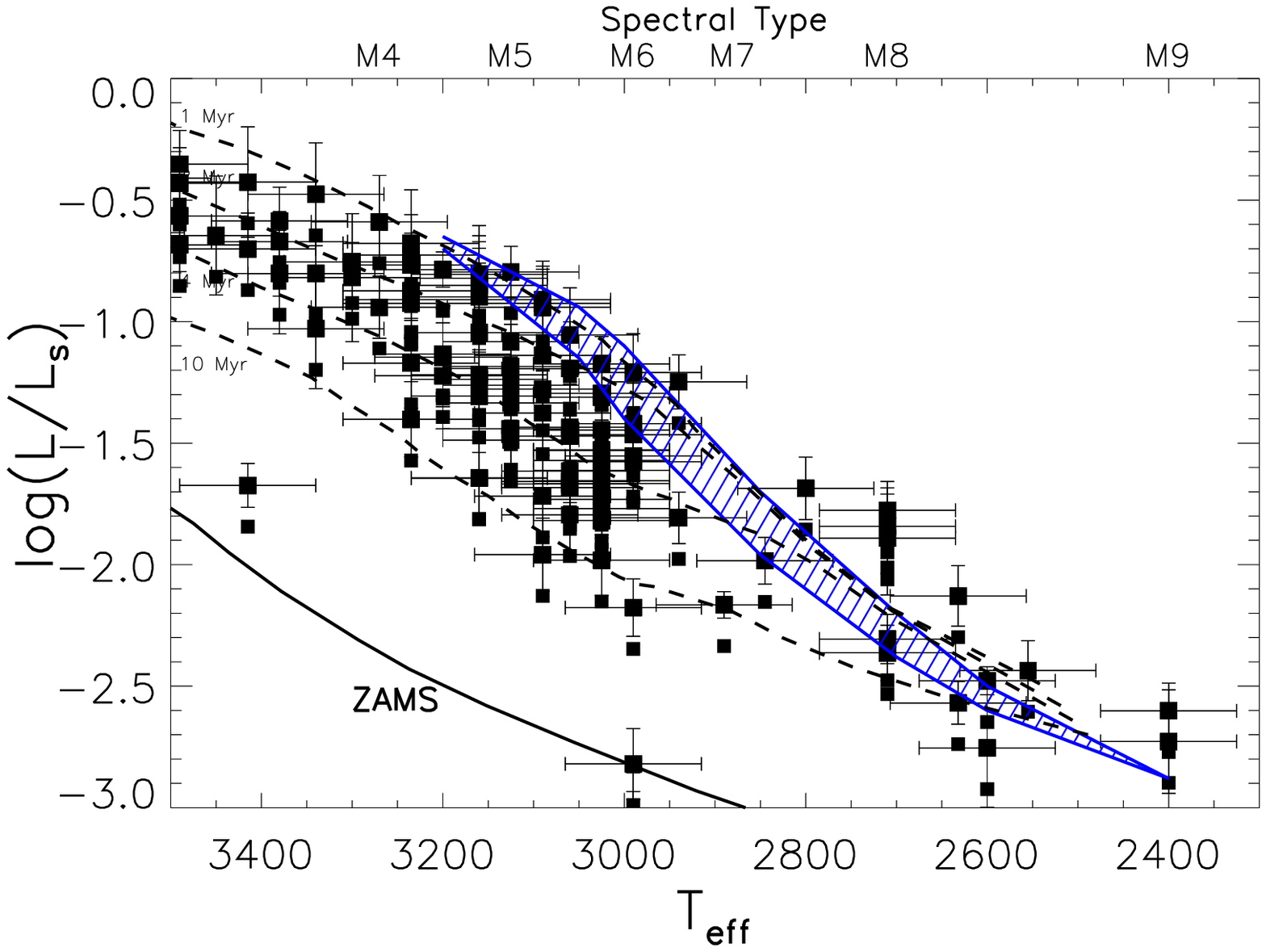}{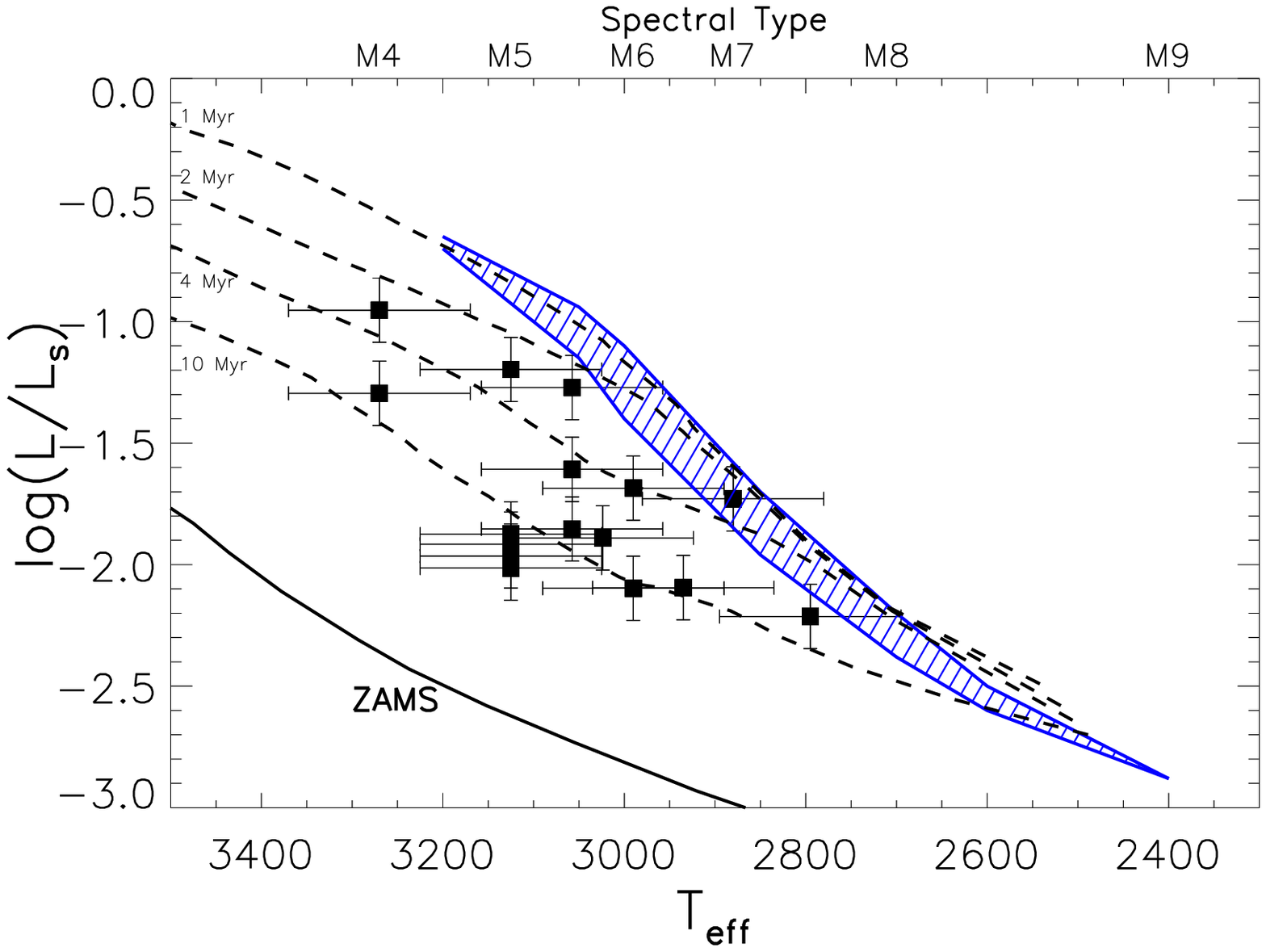}
\end{center}
\caption[]{\label{HRdiag} We plot young cluster members on the H-R
diagram, along with the deuterium-burning instability strip from PB05 (blue dashed region). 
Isochrones are from \citet{2003A&A...402..701B}, and $L_S$ is the solar luminosity. 
At upper left is $\sigma$~Orionis, with an assumed distance of 440~pc; at upper right
is Cha~I ($d\sim 160$~pc). At lower left is IC~348, with an assumed distance of
316~pc; at lower right is Upper Scorpius ($d\sim 145$~pc).
Many of the observed objects appear to lie directly on the strip, and many others have a non-negligible probability
of lying on it given their H-R diagram errors.}
\end{figure*}

In $\sigma$~Orionis, our search for signs of deuterium burning oscillations among members
was carried out with the CTIO~1.0m and {\em Spitzer} telescopes, as
detailed in Cody \& Hillenbrand (2010) and Cody \& Hillenbrand (2011),
respectively. Within the uncertainties of cluster membership
verification, there were approximately 40 objects with masses less
than $\sim$0.1~$M_\odot$. Of these, of order 15 had temperatures and
luminosities overlapping the instability strip. 
We present the H-R diagram of observed σ Ori objects with available spectral types,
for a distance of 440~pc, in Figure~\ref{HRdiag}.

Our failure to detect short-period variability in $\sigma$~Ori is inconsistent with the 
predictions of PB05, a conclusion we further quantify in the next section.
It is also at odds with previous reports of short-period variability in young 
brown dwarfs observed by \citet{2001A&A...367..218B} and \citet{2003A&A...408..663Z}.

Prior evidence for short-timescale variability was weaker (or non-existent) in the other three clusters. 
To assess prospects for pulsation here, we performed the same H-R diagram evaluation as done for $\sigma$~Ori.
In Cha~I, four objects in our sample fall squarely on the instability strip when we convert spectral types to 
temperatures using the scale of \citet{2003ApJ...593.1093L}. A further 24 have spectral types later than M4 
and therefore may be burning deuterium and subject to pulsation. We plot the full sample of Cha~I members 
in our field on the H-R diagram in Figure~\ref{HRdiag}. 

To estimate the number of our IC~348 targets that should be susceptible to the D-burning instability, we have 
plotted their positions on the H-R diagrams in Figure~\ref{HRdiag}, assuming a distance of 316~pc. Many 
objects overlap the predicted D-burning pulsation strip.
We also present the H-R diagram of observed USco objects in Figure~\ref{HRdiag}.
Of the 17 very low mass USco members observed, one lies squarely on the instability strip: 
SCH~J16173105-20504715. However, its periodogram does not exhibit any significant signals in the region 
expected for pulsation. 

A number of other targets could lie on the strip if their 1--$\sigma$ temperature 
and luminosity errors are considered, and this is true for each of the clusters examined. 
Below, we will use these uncertainties to assess the statistic likelihood of failing to observe pulsation in 
the overall monitoring dataset.

\subsection{Statistical evaluation of pulsation probabilities}

We uncovered many cases of periodic variability in the collected time series, over a wide range of 
timescales. Our detection of both rotation on $\sim$1--3 day timescales in young cluster members 
and variability on hour timescales in background field pulsators and eclipsing binaries shows that our period 
detection algorithms are robust. Yet in the search for deuterium-burning pulsation, the data 
unanimously point to one conclusion: this instability is not present in young BDs and VLMSs above 
an amplitude of several millimagnitudes in the $I$ band.

One might argue that objects in our dataset simply do not exhibit pulsation because they are not 
situated on the H-R diagram instability strip. However, the large sample size makes this possibility 
highly unlikely. To show how improbable the chances are that {\em none} of our sample have H-R diagram 
positions overlapping the instability strip, we consider temperature-luminosity probability 
distributions for each object. We take these to be two-dimensional asymmetric Gaussians, normalized and 
centered at the adopted luminosities and temperatures. The Gaussian widths are given by the associated 
1-$\sigma$ uncertainties, which are shown in the H-R diagrams in Figure~\ref{HRdiag}. The 
position of each target then corresponds to a probability that it is susceptible to pulsation, which we 
determine by integrating its distribution over the entire region of the instability strip.  For objects 
on or very close to the strip, this value is at least $\sim$20--25\%, whereas for the higher mass stars 
far from the strip it is close to zero. The probability that the position of a given object does {\em 
not} overlap with the instability strip is then 1.0 minus this quantity. The product of these values 
over all targets provides an estimate of the chance that no pulsators would be present in our sample.

We have performed this exercise for each of the clusters observed, and for alternate distances in cases 
where there is more than one possible value (IC~348 and
$\sigma$~Ori). Since the instability strip runs nearly along
isochrones and varies slowly in luminosity, changes in the adopted
distance can have a significant impact on the number of objects
expected to pulsate. We take all possibilities into consideration.

In Cha~I, we determined an 
expectation value of 3--4 objects on the strip and find a probability of 0.015 that {\em no} object 
positions actually overlap it. Turning this number around, there is a nearly 99\% chance that at least 
one object should exhibit pulsation based on its position within the instability strip, assuming that 
the theoretical calculations underpinning it (PB05) do not suffer from gross systematic errors.
Likewise in Cody \& Hillenbrand (2011), we found a 70--80\% chance that at least one of our $\sigma$~Ori 
targets observed in the infrared with {\em Spitzer} should exhibit pulsation. This is in contrast to
the lack of short period variability in that dataset. 

USco does not have many targets overlapping the instability strip, and therefore the expectation is for only 
1 or 2 objects to lie directly on it. In this region, we find a non-negligible probability of 0.22 that our 
sample did not include any pulsation candidates. For IC~348, on the other hand, we expect $\sim$11 objects on 
the strip and find a probability of 4$\times$10$^{-6}$ that none are
actually on it. This is assuming a distance of 316~pc
\citet{Herbig:4p7361}.  If we instead assume a lower distance of
260~pc based on Hipparcos parallaxes \citep{Scholz:1999p7386}, then the expectation is similar: nine objects 
on the strip and a probability of 5$\times$10$^{-5}$ that none are on it. 

Finally, we have computed probabilities for $\sigma$~Ori. The precise number of pulsation candidates depends on 
the adopted distance, which is debated to be either 350$^{+120}_{-90}$~pc, based on the Hipparcos parallax of 
$\sigma$~Ori AB, or 440$^{+30}_{-30}$ from main sequence fitting \citep{2008AJ....135.1616S}. Assuming a 
cluster distance of 440~pc \citep{2008AJ....135.1616S} we find that at least 4 targets are expected to be on 
the strip, with at most a 0.02 chance that none are. Substituting the alternate distance of 350~pc, we find 
nearly the same values (3, 0.06). The probabilities are upper limits since we do not have spectral types for 
part of the $\sigma$~Ori sample and hence cannot reliably place these objects on the H-R diagram.

\begin{figure*}
\epsscale{1.0}
\plottwo{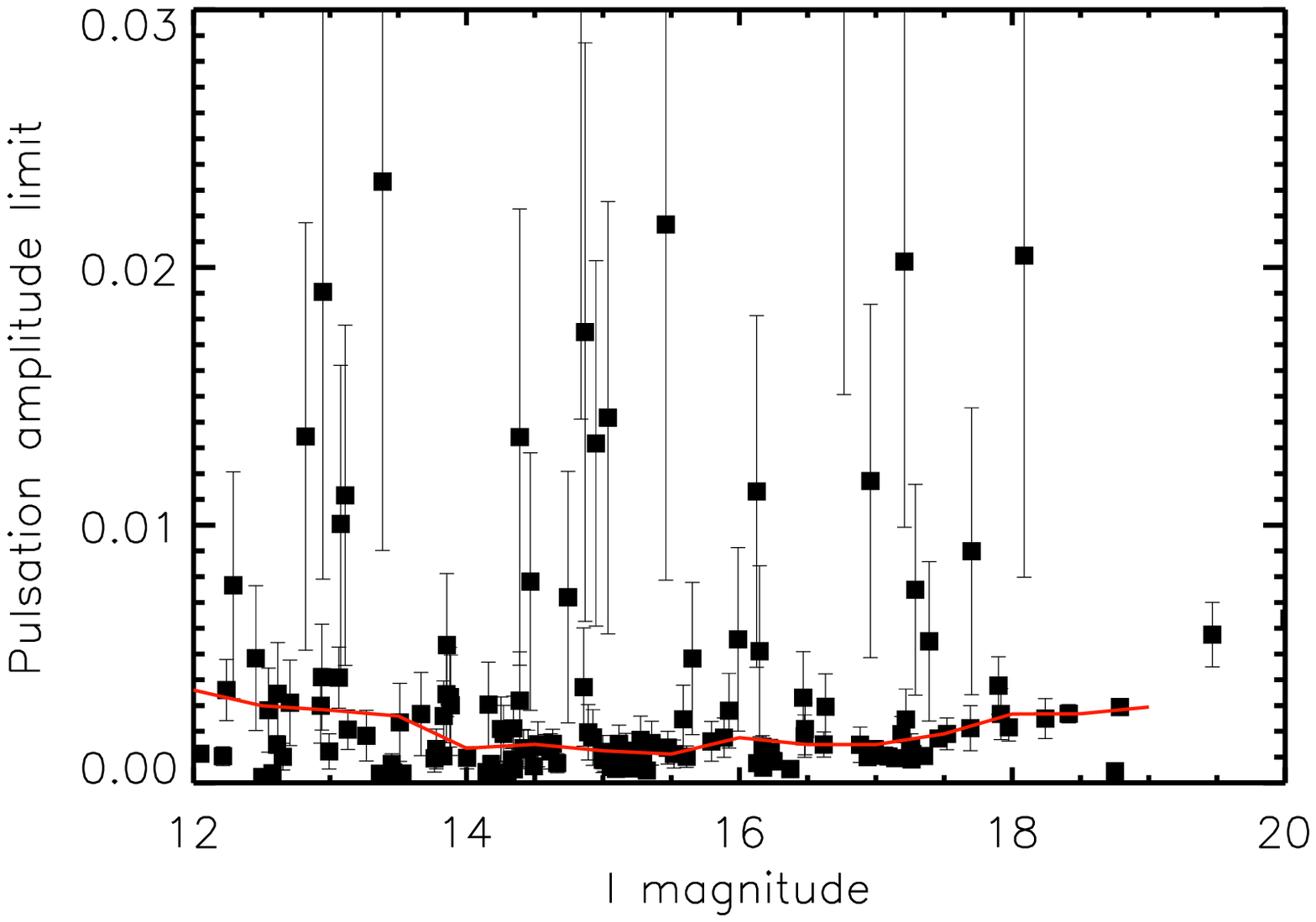}{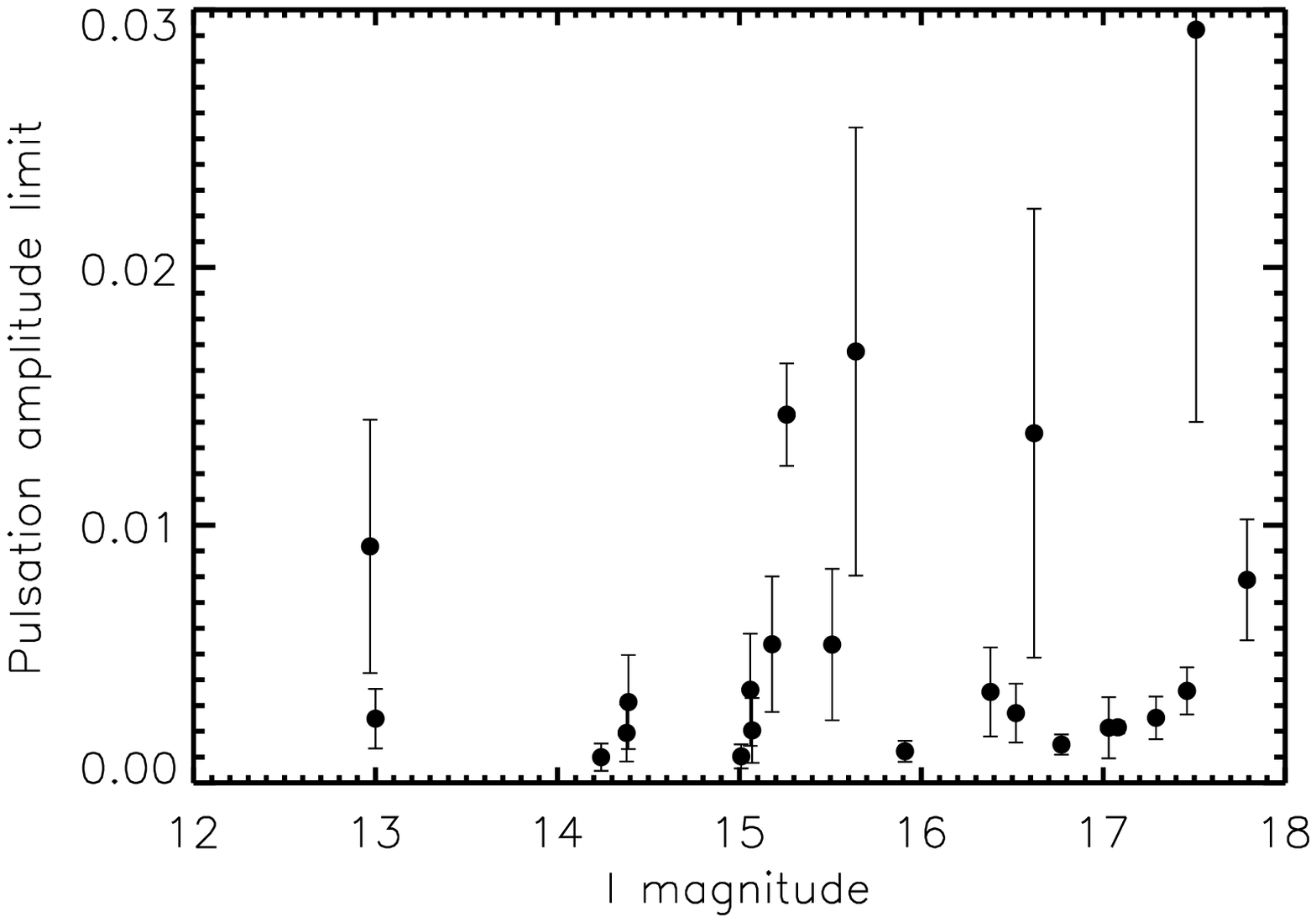}
\plottwo{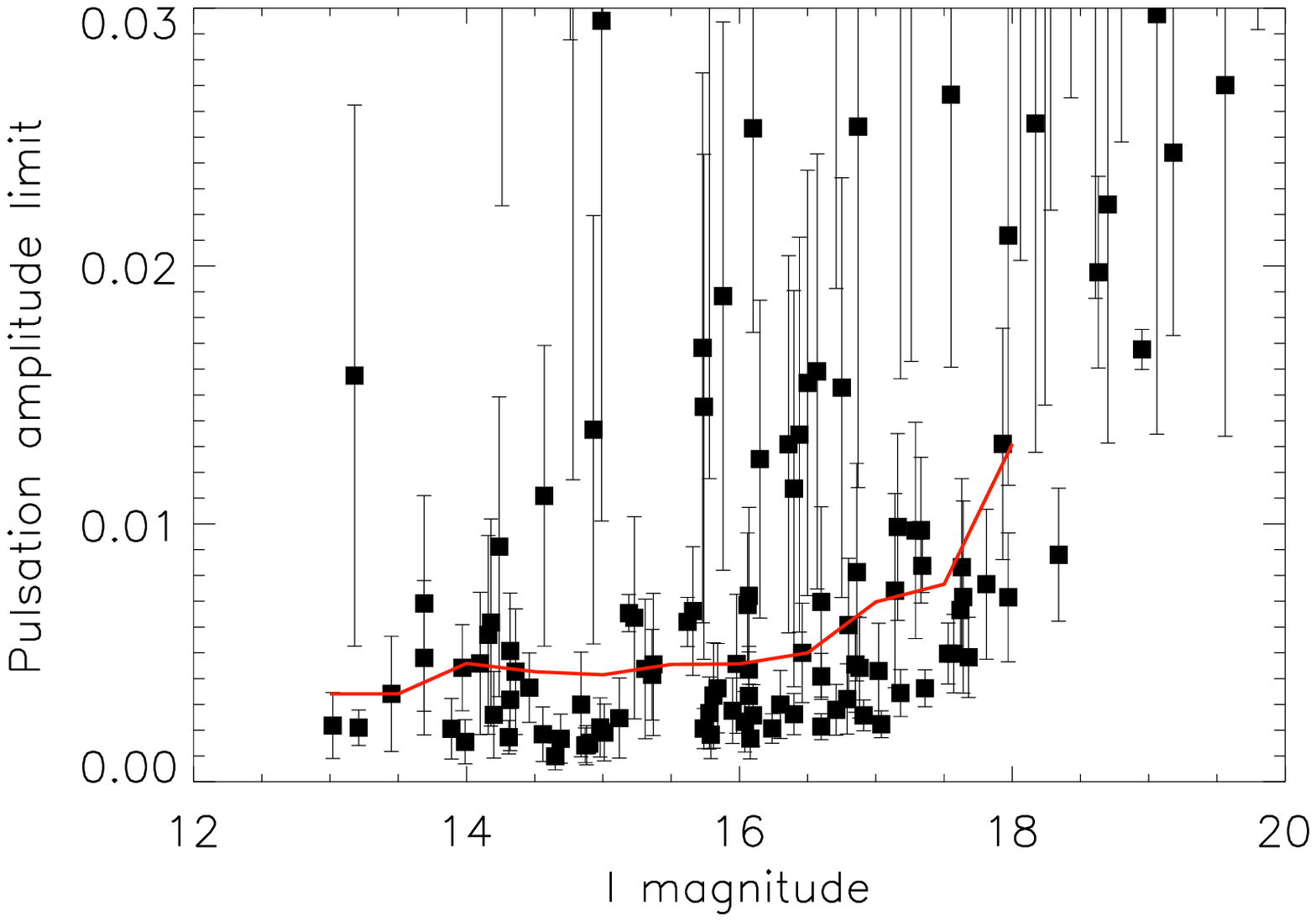}{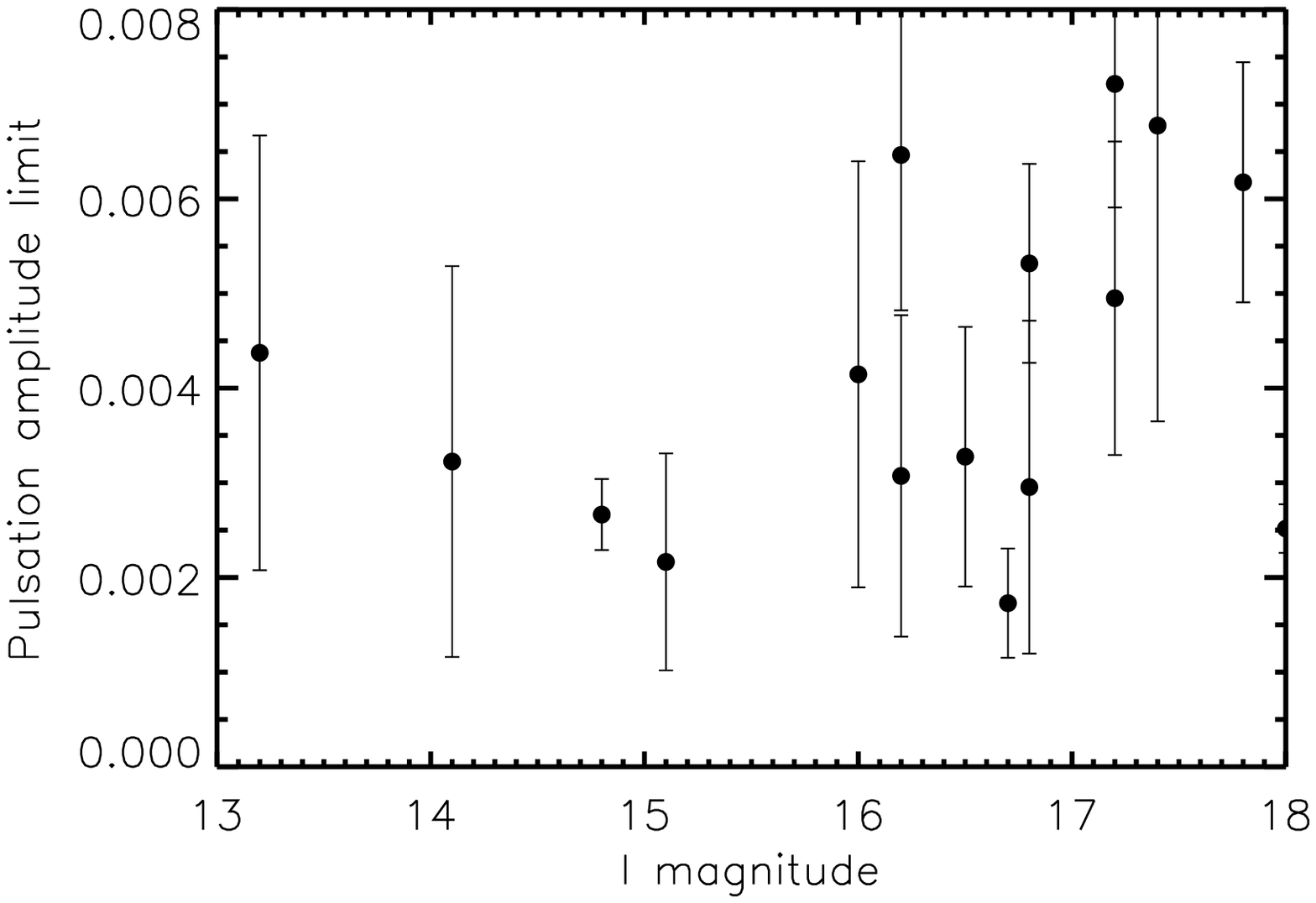}
\caption[]{\label{money}Pulsation detection limits for individual objects versus their magnitudes.
The clusters represented, from clockwise top left: $\sigma$~Ori,
Chamaeleon~I, IC~348, and Upper Scorpius.
For $\sigma$~Ori and IC~348, we have drawn a median curve, binned by magnitude, in red. Based on the position of the 
instability strip, we would expect pulsating objects to have magnitudes of $\sim$14 and fainter.
While in general the limits are quite low--in the millimagnitude range--there is a large population of outliers in 
which high-amplitude intrinsic aperiodic introduced significant power into the higher frequency regions of the 
periodogram.} 
\end{figure*}

In conclusion, we expect with high confidence to have observed deuterium-burning oscillations if it is
present at observable amplitudes. We now quantify the overall detection limits by considering the power-law 
fit to the periodograms of each observed young cluster member. These curves, of form $A/(f+B)+C$ for frequency 
$f$ and constants $A$, $B$, and $C$, trace out the noise level as a
function of frequency (see Figure~\ref{periodos}). For each object analyzed, we take the fit 
values at 5~cd$^{-1}$ ($\sim$5 hours) and 25~cd$^{-1}$ ($\sim$1 hour) as representative of the 1--$\sigma$
level above which no pulsation is observed. We display these values as a function of object magnitude 
in Figure~\ref{money} to illustrate the collective limit imposed by our entire dataset. 

The median amplitude limit is several millimagnitudes.  Objects with high-amplitude aperiodic variability 
(see the Appendix) are exceptions, as they have excess periodogram noise which is intrinsic. The rest of 
our targets, however, have maximum amplitudes in the periodogram of at most 0.002 to 0.004 magnitudes. This 
represents the threshold above which we detect no periodicities. We conclude that if deuterium-burning 
pulsation is present in any of our sources, then its amplitude must be below this level.

\section{Conclusion}

Despite exquisite photometric sensitivity, we have not detected signs of short-period variations in any of 
our young BD and VLMS targets. Although the theory of PB05 does not preclude very low amplitudes, we 
suspect that the failure to find pulsation is indicative of a physical damping mechanism operating within 
these objects. The convective timescale is over two orders of magnitude longer than the pulsation 
timescale, but it becomes quite short near the (sub)stellar surface layers. Neglect of the energy 
exchange between pulsation and convection may have led to overly optimistic predictions of mode amplitude 
growth. Indeed, models of this interaction in other types of stars have recently shown that convection can 
quench pulsation under some circumstances \citep{Gastine:2010p2083}.

To continue the search for pulsation and probe to lower amplitudes, future campaigns will need to 
produce extraordinarily high precision photometry. Data of this quality is available through the 
{\em CoRoT} mission, but only for stars brighter than $\sim$15th
magnitude in the optical in the NGC~2264 cluster. K2, the successor
mission to {\em Kepler} shows further promise as it will provide
exquisite monitoring on hundreds of Upper Scorpius members. 

\acknowledgements{Based on observations made with the NASA/ESA Hubble Space Telescope, obtained at the Space Telescope Science Institute, which 
is operated by the Association of Universities for Research in Astronomy, Inc., under NASA contract NAS 5-26555. These observations are 
associated with program 11610. A.M.C. thanks the CTIO Telescope Operations staff for help in carrying out observations. Observation time on 
SMARTS consortium facilities was awarded through the National Optical Astronomy Observatory, operated by the Association of Universities for 
Research in Astronomy, under contract with the National Science Foundation.}

\clearpage

\section*{Appendix}

\section{Contamination by aperiodic variability}

In searching for periodicities among our young cluster targets, we encountered a number of light curves with
RMS values well above the median for field objects of similar magnitude. This aperiodic variability introduces low-frequency
noise into the periodograms and raises the threshold over which we can detect pulsation signals. While this 
behavior may ``contaminate'' the pulsation search, it is interesting in its own right, as it offers insights into
accretion and disk properties. We have assembled a collection of aperiodic variables in each of the clusters apart from
USco, which did not have a sufficient sample size to mine non-periodic variables. 

In general, we used plots of light curve RMS versus magnitude (e.g., Figures \ref{apermagrms} and \ref{apermagrms1}) to select objects with 
variability at the 99\% confidence level (see CH10 Section~6 for a full description of detection procedures). In $\sigma$~Ori we uncovered 42 
aperiodic variables with amplitudes from a few percent up to a full magnitude. Combining these with the periodically variable objects
in that cluster, we found a total variable fraction of 69\% for $\sigma$~Ori members (CH10).

\begin{figure*}
\begin{center}
\includegraphics[scale=0.7]{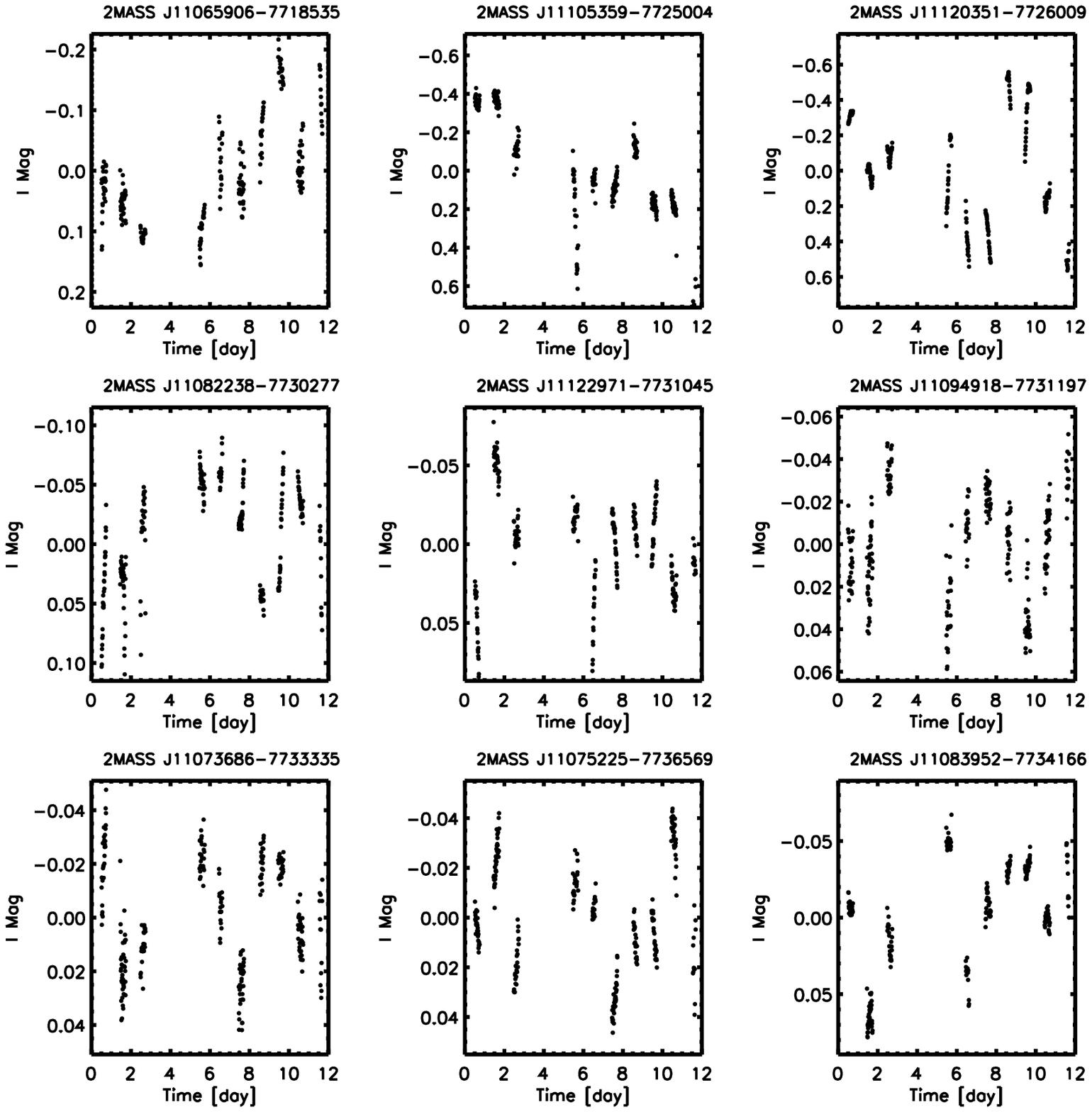}
\includegraphics[scale=0.7]{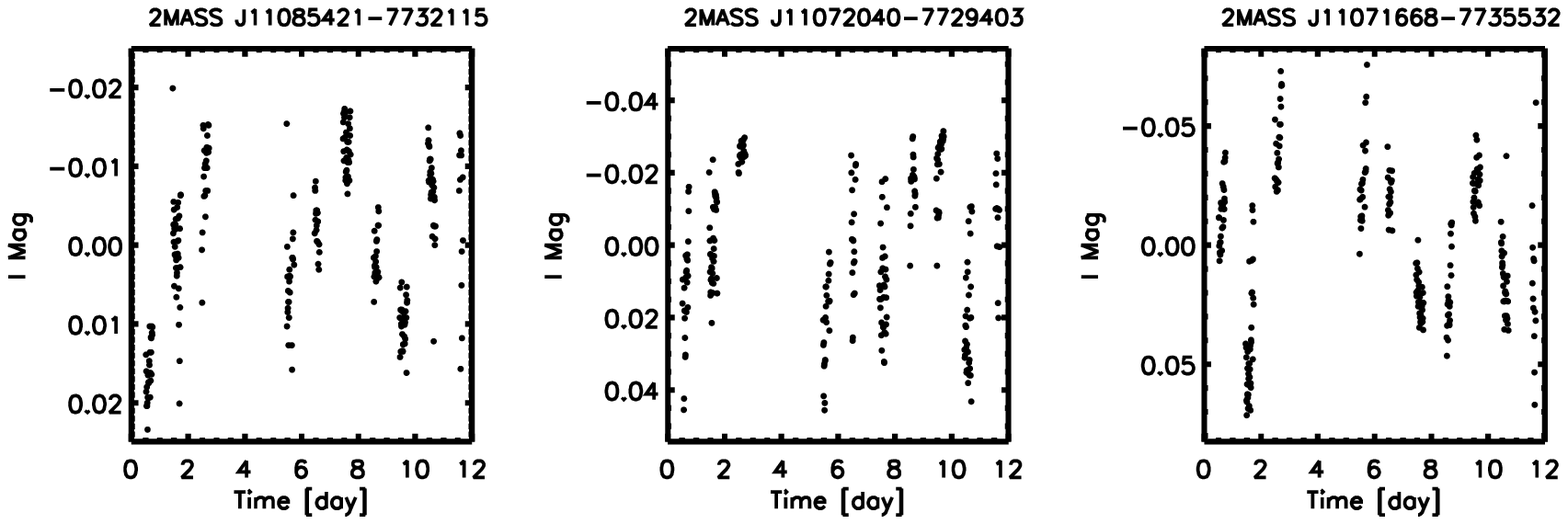}
\end{center}
\vspace{-8.3cm}
\caption[]{\label{chaIaperlightcurves}Light curves of very-low-mass
Cha~I members selected as aperiodic based on large $\chi^2$
values and lack of periodicities. Membership information is available in Table~6; 2MASS J11122971-7731045 is a new Cha~I candidate.}
\end{figure*}

Using the same detection approach in the Cha~I field, we found 13 aperiodic variables, all but one of which are confirmed cluster members. The 
remaining object, 2MASS~J11122971-7731045, has peak-to-peak variations of just over 0.1 magnitudes in the $I$ band. We provide the list of aperiodic 
variables in Table~6, and show their RMS values in Figure~\ref{apermagrms}. Their light curves are provided in Figure~\ref{chaIaperlightcurves}. Of 
note, three stars previously listed as periodic \citep{Joergens:2003p3530} appear to be aperiodic, since they are detected as variable based on light 
curve RMS but do not show distinct signals in their periodograms. These are 2MASS~J11085421-7732115 (CHXR~78C), 2MASS~J11075225-7736569 
(Cha~H$\alpha$~3), and 2MASS~J11083952-7734166 (Cha~H$\alpha$~6).

In the Cha~I sample, the total variability fraction (including periodic objects) is between 53 and 69\%, depending on the membership status 
of several newly identified variable stars. This is statistically comparable to the fraction found in $\sigma$~Ori (CH10). The Cha~I variability 
classification is divided into roughly equal proportions of periodic and aperiodic objects. Among the eight variables with no prior 
membership information, five have light curves and colors characteristic of field eclipsing binaries or pulsators. Three may be new members, 
and we note these in Table~6.

\begin{figure*}
\begin{center}
\includegraphics[scale=0.7]{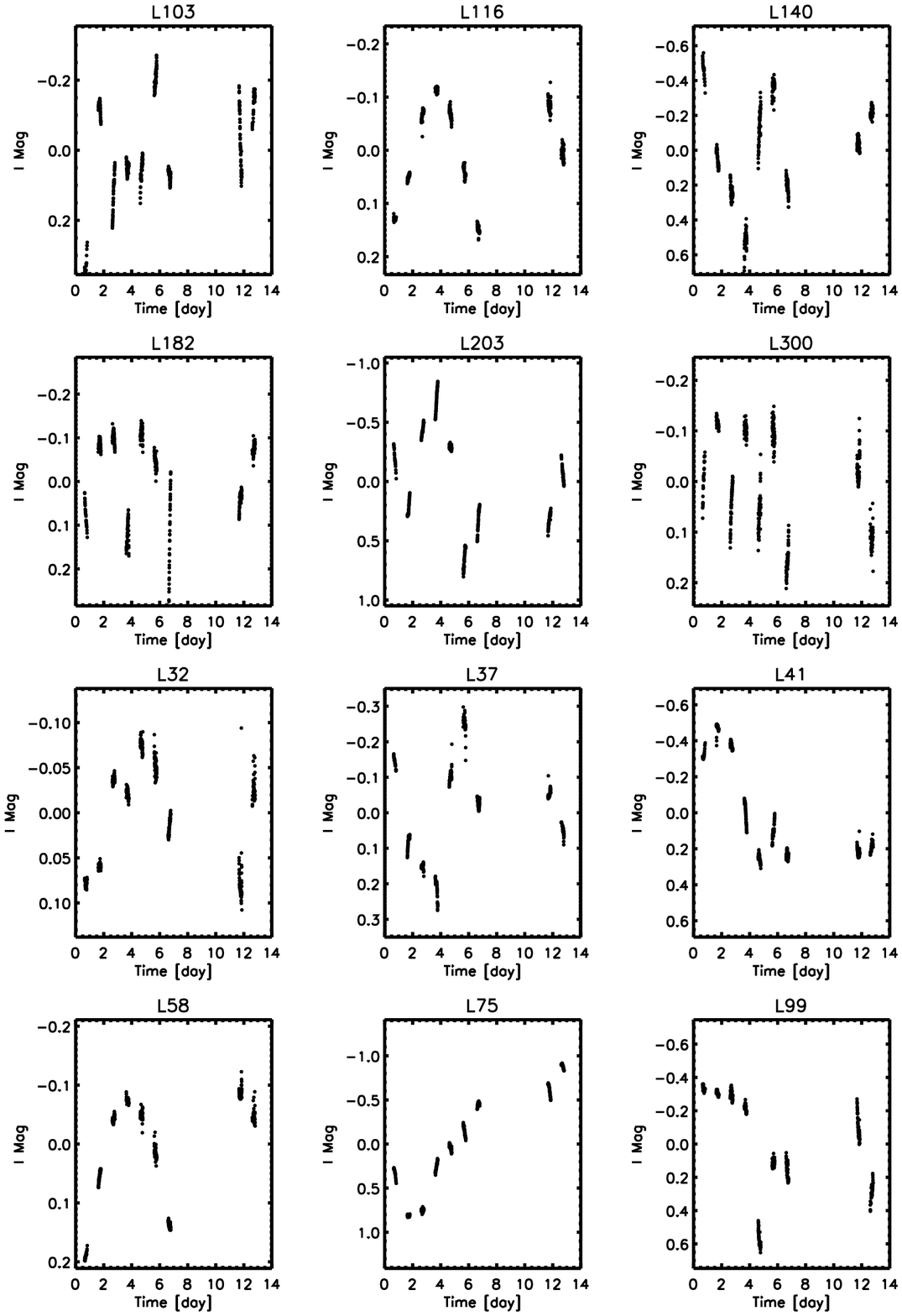}
\end{center}
\caption[]{\label{IC348aperlightcurves}Light curves of IC~348 members selected by eye as being prominent variables.
Ids are as in Table 3.}
\end{figure*}

In the IC~348 sample, there were no non-variable field stars available for determination of the photometric uncertainty as
a function of magnitude. As a result, we could not calculate accurate $\chi^2$ values for the light curves from this cluster.
However, a rough cut-off for RMS values indicative of variability (i.e., upper line in Figure~\ref{apermagrms1}) suggests that 
there are a number high-amplitude variables. We examined the light curves of objects above the upper curve in Figure~\ref{apermagrms1}
and removed from consideration those for which artificial systematic effects (i.e., tracking and flatfielding errors) appeared to 
cause a high RMS. We are left with nine IC~348 objects with strong variability; these light curves are depicted in 
Figure~\ref{IC348aperlightcurves}. Their identification numbers, from \citet{Luhman:2003p2777}, are provided above each panel.

Beyond identification of the aperiodic variables in our dataset, we can assess correlations between their flux behavior and the presence of 
circumstellar disks. We cross-referenced our photometric samples in $\sigma$~Ori and Cha~I with those of \citet{2008ApJ...688..362L}, 
\citet{Luhman:2008p2378}, and \citet{Luhman:2008p2387}, which provide {\em Spitzer} Infrared Array Camera (IRAC; 3.6--8.0 $\mu$m) and in some 
cases Multiband Imaging Photometer for Spitzer (MIPS; 24 $\mu$m) photometry. In $\sigma$~Ori, we find that 133 of 153 confirmed or candidate 
members in our time series dataset have {\em Spitzer} photometry, including 57 of 65 cluster periodic variables. For the Cha~I sample, all 37 
cluster members monitored in the $i$ band have available IRAC measurements, and in many cases, MIPS data in addition. While there is {\em 
Spitzer} data available for the IC~348 cluster as well, we have not included it in the analysis here since extensive comparison of 
photometric periods with infrared excess was already carried out by \citet{2006ApJ...649..862C}. Furthermore, our own IC~348 photometry is 
difficult to cleanly separate into the periodic and aperiodic categories due to the systematics resulting from the lack of tracking on the P60 
telescope.

\begin{figure}
\begin{center}
\includegraphics[scale=0.5]{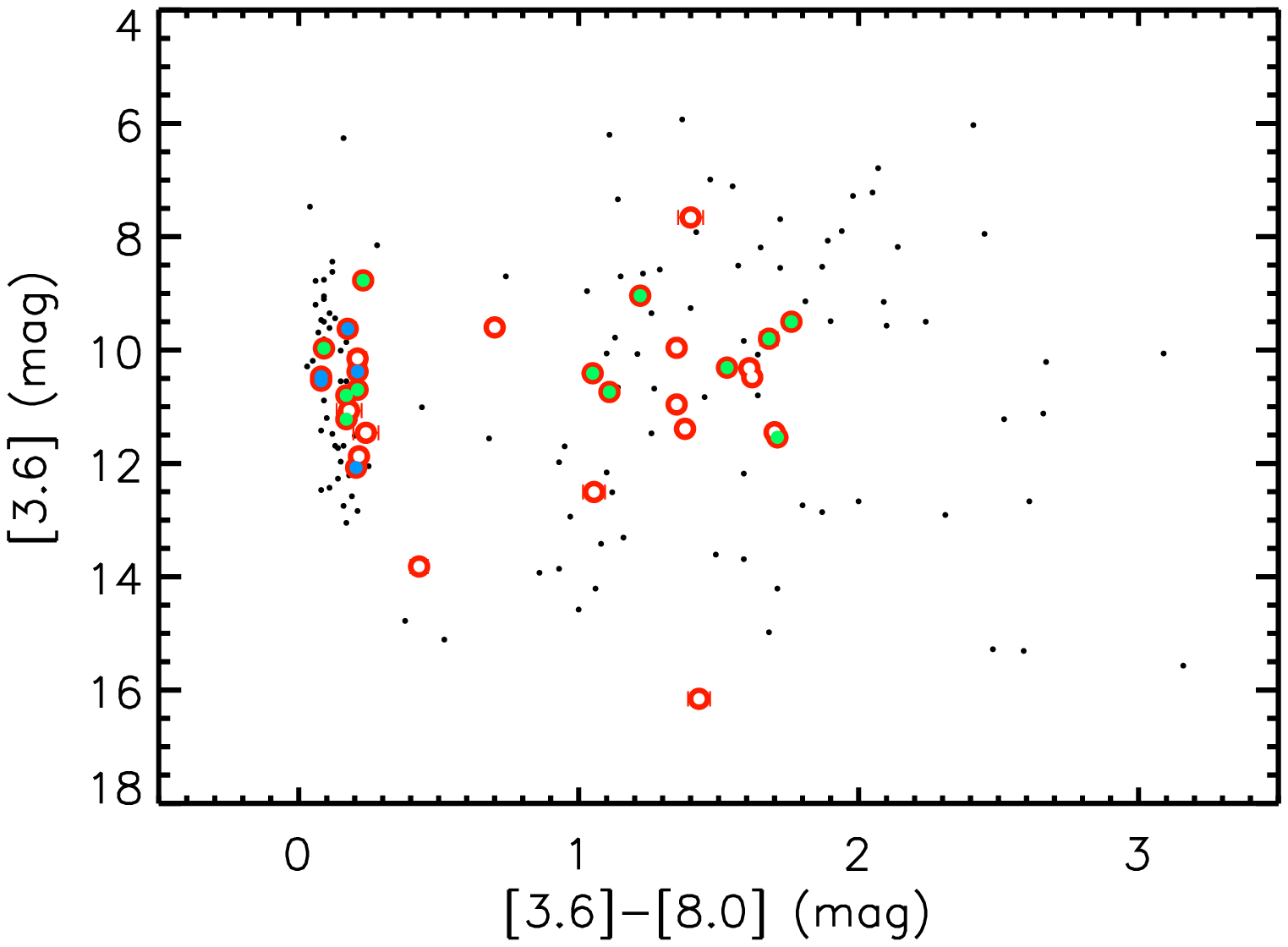}
\end{center}
\caption[]{\label{spitcolmag}{\em Spitzer} photometry of likely Cha~I members. 
Objects found in our photometric sample are marked with red circles, while those out of the fields are left as dots. 
Error bars are included but in many cases too small to see. Aperiodic variables detected in our photometry are 
overplotted as filled green circles, while periodic variables in 
our sample are marked by filled blue circles. The nearly vertical cluster of objects near [3.6]-[8.0$]=0$ is the 
sequence of colors and magnitudes pertaining to bare photospheres.}
\end{figure}

We display in Figure~\ref{spitcolmag} the distribution of $Spitzer$/IRAC 3.6--8.0~$\mu$m colors for all objects in our Cha~I dataset with 
available infrared photometry; the equivalent plot for $\sigma$~Orionis is shown in Figure~13 of CH10. As seen in these diagrams, the samples 
split relatively cleanly into two groups, with the narrower blue sequence near [3.6]-[8.0$]=0$ representing bare photosphere colors. The 
cloud of objects with [3.6]-[8.0] colors between 1 and 2 is indicative of infrared excesses signifying the presence of a dusty disk. While 
the sequence of {\em photospheric} colors is fairly well defined, several ambiguous objects lie between 0.3 and 0.7 magnitudes. We have 
therefore chosen a somewhat conservative disk selection criteria of [3.6]-[8.0$]>0.7$ \citep[e.g.,][]{2007ApJ...671..605C} so as to omit 
these objects from the disk sample.

In total, we identify 21 disk-bearing Cha~I members with both photometry from our campaign and {\em Spitzer} colors 
indicative of disks. The resulting disk fraction in our sample is 57$\pm$9\%. We find that our disk 
identification is entirely consistent with that of \citet{Luhman:2008p2378} and \citet{Luhman:2008p2387} (based on the 
same {\em Spitzer} data). The full listing of disk classifications is provided in Table~6.

In Figure~\ref{spitcolmag} we have distinguished variable objects (blue for periodic; green for aperiodic) from the non-variables 
(white) in the {\em Spitzer}/IRAC 
color-magnitude diagram. In the small Cha~I sample, none of the periodic variables has an infrared excess suggestive of 
a disk. As was the case with $\sigma$~Orionis (CH10), we can associate disks with the majority of aperiodic variables in 
our sample and lack of a disk with most of the periodic variables. But a number of objects do not fit these scenarios. 

\begin{figure}
\begin{center}
\includegraphics[scale=0.5]{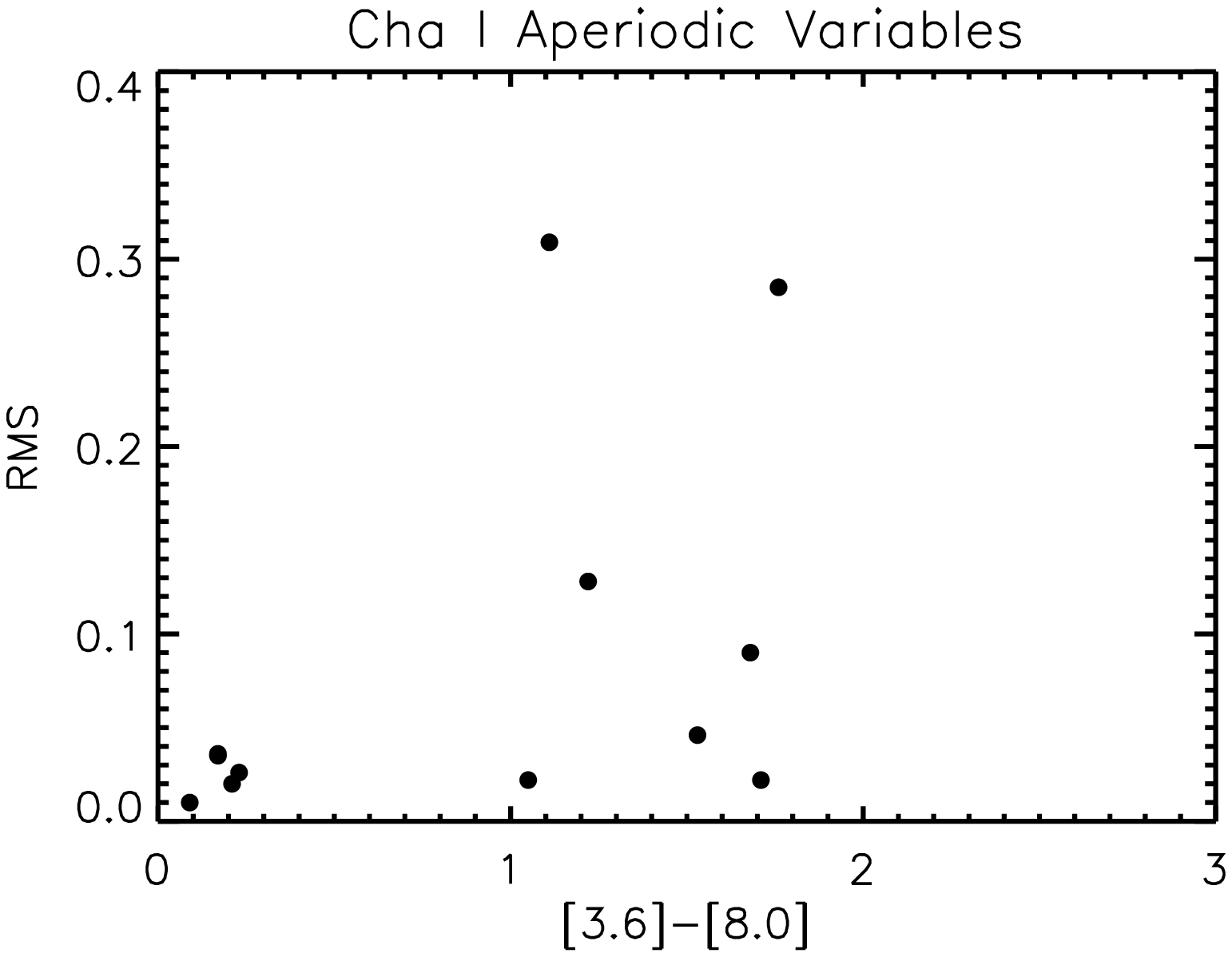}
\end{center}
\caption[]{\label{spitrms}{\em Spitzer} [3.6]-[8.0] color versus light curve $I$-band RMS value for our aperiodic variables
in Cha~I.}
\end{figure}

Five Cha~I members display aperiodic variability but no sign of infrared excess in the {\em Spitzer} data.
This curious small population of objects with RMS values ($\sim$0.01--0.03 magnitudes) much 
lower than the other aperiodic variables and have H$\alpha$ pEW and [3.6]-[8.0] values suggesting {\em absence} 
of accretion or an associated disk. In addition to light curves in which variability is clearly obvious 
by eye, these objects have $\chi^2$ values high enough that their status as variables is not in doubt. All but 
one have $\chi^2>4.5$, or less than $10^{-5}$ probability that the light curve trends arose by chance; the 
remaining object (2MASS J05383922-0253084) has a $\chi^2$ value of 2.85, or an estimated 0.4\% probability 
that its light curve behavior is explained by noise. We show in Figure~\ref{spitrms} the RMS and infrared 
colors for Cha~I members; Figure 15 of CH10 illustrates another such cluster of low-amplitude aperiodic variables
with no signs of circumstellar disks.

A similar phenomenon was observed in the IC~348 cluster, in which a number of weak T~Tauri stars 
(i.e., weak H$\alpha$) were found to be erratic variables by \citet{2005MNRAS.358..341L}. These results 
bring into question our ability to determine which cluster members are truly surrounded by disk material, 
which ultimately affects the analysis of rotation and possible disk locking. It appears from these light 
curves that a percentage of young objects retain enough gas and/or dust beyond the time that we would 
expect their disks to be fully cleared based on infrared observations. Alternatively, we may be viewing rapid 
evolution of magnetic spot features on the stellar surface.

\newpage
\bibliographystyle{apj}
\bibliography{pulsation}

\end{document}